\documentclass{article}

\PassOptionsToPackage{numbers, compress}{natbib}

\usepackage[main, final]{neurips_2026}
\usepackage{kim-paper-macros}
\usepackage{wrapfig}
\usepackage{needspace}

\usepackage[utf8]{inputenc} 
\usepackage[T1]{fontenc}    
\usepackage{graphicx}
\usepackage{float}
\usepackage{capt-of}
\usepackage{hyperref}       
\usepackage{url}            
\usepackage{booktabs}       
\usepackage{amsfonts}       
\usepackage{nicefrac}       
\usepackage{microtype}      
\usepackage{xcolor}         
\usepackage{colortbl}       
\graphicspath{{figures/}}

\newcommand{\sys}{Dooly}

\newcommand{\mconfig}{\textnormal{\textsc{model\_config}}}
\newcommand{\numtoks}{\textnormal{\textsc{num\_toks}}}
\newcommand{\numreqs}{\textnormal{\textsc{num\_reqs}}}
\newcommand{\mix}{\textnormal{\textsc{mix}}}

\title{\sys: Configuration-Agnostic, Redundancy-Aware Profiling for LLM Inference Simulation}

\makeatletter
\author{%
  Joon Ha Kim \quad Geon-Woo Kim \quad Anoop Rachakonda \quad Daehyeok Kim \\
  \texttt{\{jhk.james1110, gwkim, anooprac, daehyeok\}@utexas.edu} \\
  The University of Texas at Austin
}
\makeatother

\begin{document}

\maketitle
\begin{abstract}
Selecting the optimal LLM inference configuration requires evaluation across hardware, serving engines, attention backends, and model architectures, since no single choice performs best across all workloads.
Profile-based simulators are the standard tool, yet they hardcode their operation set to a specific configuration and re-profile every operation from scratch, making exploration prohibitively expensive.
This cost stems from a missing structural understanding: every input dimension of each operation is fixed by the model configuration or determined by the incoming request.
Many model-configuration values (\eg head size, layer count) recur across models, so the same operation runs in many configurations; a single sweep over the request-dependent dimensions can serve them all.
We present \sys{}, which exploits this structure to achieve \emph{configuration-agnostic, redundancy-aware} profiling.
\sys{} performs a single inference pass, labels each input dimension with its origin via taint propagation, and selectively profiles only operations absent from its latency database; stateful operations such as attention are isolated by reusing the serving engine's own initialization code, eliminating manual instrumentation.
From this database, it builds latency regression models that serve as a drop-in backend for existing simulators.
Across two GPU platforms, three attention backends, and diverse model architectures, \sys{} achieves simulation accuracy within 5\% MAPE for TTFT and 8\% for TPOT while reducing profiling GPU-hours across 12 models by 56.4\% compared to the existing profiling approach.
We have open-sourced \sys{} at \url{https://dooly-project.github.io}.
\end{abstract}

\section{Introduction}
\label{sec:intro}

Deploying an LLM inference system requires navigating a configuration space spanning hardware, serving engines~\cite{vllm,sglang}, attention backends~\cite{flashinfer,flashattention}, model architectures, and workload characteristics.
No single configuration performs optimally across this space: the best (model, backend) pair can invert across workload types and sequence lengths, making pre-deployment evaluation indispensable (\autoref{sec:configurations}).
Profile-based simulators~\citep{vidur, revati, llmservingsim, llmservingsim2.0, apex, frontier} have become the standard tool for this evaluation, collecting GPU execution latencies to predict end-to-end serving performance without running a live system.

Today's simulators, however, suffer from two limitations that hinder this exploration.
First, they hardcode the operation set and the inputs to each operation, locking each simulator to a specific configuration; any stack update or new architecture requires manual rework to restore coverage.
Second, because simulators cannot recognize when two configurations invoke the same underlying computation, they re-profile each from scratch.
Profiling four similar-sized models in Vidur~\cite{vidur} takes 27.7 wall-clock hours on an A100 GPU, and this cost grows linearly with each new model or backend.

Neither limitation is inherent to profiling; both stem from a missing structural understanding of LLM inference, which we develop through two observations.
First, every input dimension of each operation derives from one of two sources: model configuration parameters (hidden dimension, number of heads) or incoming request characteristics (token count, batch size).
Second, many model-configuration values recur across models, so a large fraction of operations are shared across configurations: models with the same attention geometry invoke identical operations regardless of tokenizer differences.
Thus, labeling each dimension with its origin lets a profiler detect equivalent operations across configurations and eliminate redundant profiling runs.

Building on these observations, we present \sys{}, a framework for \emph{configuration-agnostic, redundancy-aware} LLM inference profiling.
\sys{} dynamically extracts all operations invoked by a target model through a single inference pass with a dummy prompt, labeling each input dimension with its origin.
It then profiles only operations absent from its latency database: from these labels, it computes \emph{signatures} that canonically identify each operation by its model-derived dimensions, allowing it to skip equivalent operations across configurations.
From the latency database, \sys{} trains per-signature regression models that existing simulators can adopt as a drop-in profiling backend; for end-to-end simulation, \sys{} reuses the serving engine's own scheduler to reproduce exact scheduling decisions and queries these regression models as it walks the model's call graph.

Two key techniques make this practical across diverse LLMs and serving stacks.
First, \emph{taint propagation} labels each input dimension by tainting values at the serving engine's model-configuration and request-processing entry points and propagating them through the inference pass; the resulting labels let \sys{} compute operation signatures from model-derived dimensions only, enabling correct deduplication.
Second, \emph{hierarchical context resolution} handles stateful operations, such as attention kernels that require KV cache memory and request metadata, by reusing the serving engine's own initialization code to reconstruct the necessary runtime state.

We implement \sys{} for vLLM and SGLang, and evaluate it across two hardware platforms (NVIDIA A100 and H100), three attention backends (FlashInfer~\cite{flashinfer}, FlashAttention~\cite{flashattention}, TritonAttention~\cite{tritonattention}), and 12 model architectures spanning dense~\cite{gqa,mha}, MoE~\cite{moe,mixtral-of-experts}, and heterogeneous model families~\cite{sliding-window}.
\sys{} predicts vLLM's serving latency within 5\% MAPE for TTFT and 8\% for TPOT, and reproduces batch scheduling decisions within 0.5\%.
Across 12 models profiled with three attention backends, it reduces total profiling GPU-hours by 56.4\% compared to the existing profiling approach.

\section{Background and Motivation}
\label{sec:background}

\subsection{Configuration Space in LLM Inference Deployment}
\label{sec:configurations}

Deploying an LLM inference system requires navigating a large configuration space whose choices substantially affect performance, typically characterized by Time To First Token (TTFT), Time Per Output Token (TPOT), and throughput~\citep{sloserve, distserve, splitwise, sarathi-serve}.
We organize this space along four key axes:
  
\textbf{(1) Hardware (H):} GPU selection (\eg NVIDIA A100 or H100), each with distinct memory capacity and compute throughput; 
\textbf{(2) System (S):} The software runtime stack, primarily defined by the \textit{serving engine}~\citep{vllm, sglang}, which manages request batching and KV-cache memory, and the \textit{attention backend}~\citep{flashattention, flashinfer}, which provides GPU-optimized attention kernels; 
\textbf{(3) Model (M):} The model architecture, including hidden dimensions, layer counts, and structural variants such as Mixture-of-Experts (MoE); and
\textbf{(4) Workload (W):} Runtime load characteristics, including request arrival rates, prompt lengths, and batch sizes.

  \begin{wrapfigure}{r}{0.6\textwidth}
  \centering
  \makeatletter
  \vspace{-22pt}
  \includegraphics[width=0.7\linewidth]{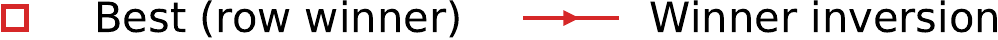}\\
  \begin{subfigure}[t]{0.5\linewidth}
      \centering
      \includegraphics[width=\linewidth]{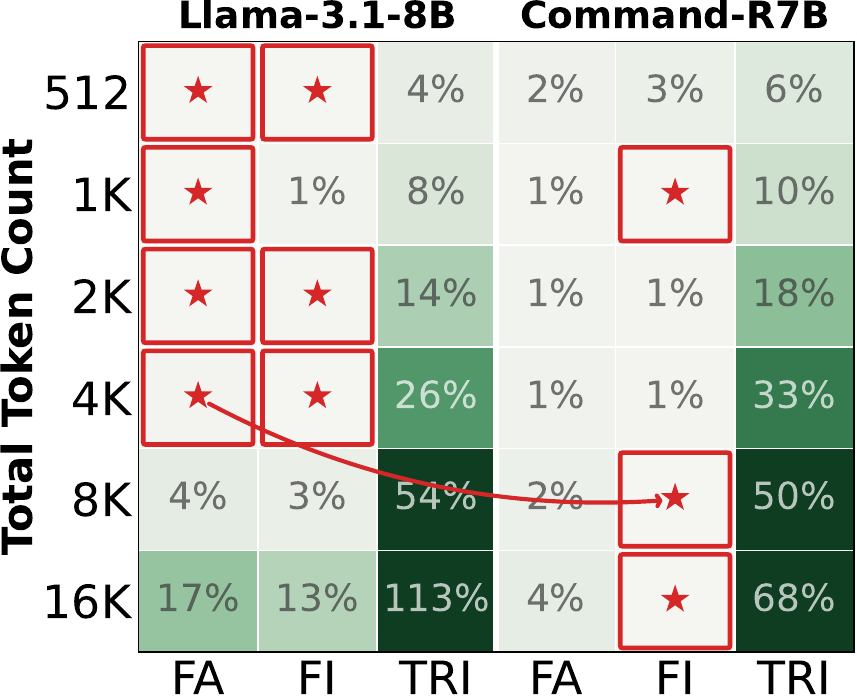}
      \vspace{-13pt}
      \caption{Prefill-Heavy Workload}
      \label{fig:configuration_space:measured_a100_prefill}
    \end{subfigure}
    \begin{subfigure}[t]{0.475\linewidth}
      \centering
      \includegraphics[width=\linewidth]{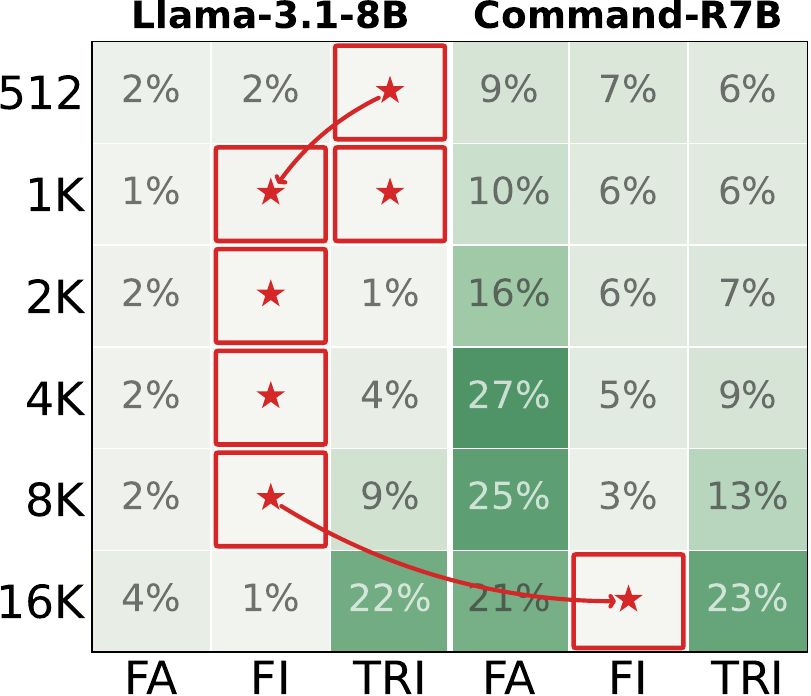}
      \vspace{-13pt}
      \caption{Decode-Heavy Workload}
      \label{fig:configuration_space:measured_a100_decode}
    \end{subfigure}
    \vspace{-5pt}
    \caption{
    Performance variation across configurations. 
    TTFT and TPOT are reported for prefill and decode workloads. 
    Red lines mark winner inversions; ties are within 0.5\% of the best
    (FA: FlashAttention, FI: FlashInfer, TRI: TritonAttention).}
    \label{fig:configuration_space}
    \vspace{-10pt}
\end{wrapfigure}
\vspace{-1.55pt}

No single configuration performs optimally across the full $H \times S \times M \times W$ space.
We demonstrate this empirically using two similar-sized models, Llama-3.1-8B~\citep{llama3.1-8b} and Command-R7B~\citep{command-r7b}, on an A100 GPU; \autoref{fig:configuration_space} reports TTFT for prefill-heavy and TPOT for decode-heavy workloads, where total token count denotes prompt size and cached context size, respectively.
Even at $\leq$1K tokens, where all (model, backend) combinations lie within 10\% of each other, the winner shifts by workload: Llama-3.1-8B with FlashAttention leads prefill-heavy workloads while Llama-3.1-8B with FlashInfer and TritonAttention leads decode-heavy ones.
Above 4K tokens, the prefill-heavy winner changes to Command-R7B, whose interleaved sliding-window attention confines most layers to a 4K context window and caps per-layer attention cost as sequences grow.

The decode-heavy winner is more stable: Llama-3.1-8B with FlashInfer leads until it, too, inverts to Command-R7B at 16K tokens.
At 16K tokens, choosing the wrong model alone, even with its best backend, costs 13\% in prefill latency; the worst (model, backend) pairing falls 113\% behind the optimum.
Backend choice within each (model, workload) pair also matters: each backend implements a distinct mix of compile-time kernel specialization and runtime kernel selection, yielding different latency profiles under identical model configurations.
The optimal deployment must thus be discovered empirically.

\subsection{Limitations of Profile-Based LLM Simulators}
\label{sec:profile_simulators}

Since exhaustively running each configuration on a live system to measure performance is impractical, profile-based simulators~\citep{vidur, revati, llmservingsim, llmservingsim2.0, apex, frontier} have become the standard tool for estimating LLM performance before deployment.
These simulators treat the GPU as a black box, collecting the execution latency of operations and modules via PyTorch's Kineto profiler~\citep{kineto}, which records GPU activity through NVIDIA's CUPTI~\citep{cupti}.
We define an \textit{operation} as a low-level CPU call that dispatches a GPU kernel (\eg \texttt{aten::linear}), and a \textit{module} as a coarser-grained PyTorch block composed of multiple operations (\eg \texttt{RowParallelLinear}).

Two limitations undermine the cross-configuration exploration these simulators are meant to enable.

\mypara{Static configuration dependencies.}
Today's simulators \emph{fix} the system and model configuration axes by hardcoding both the operation set to profile and the arguments used to invoke each operation, tying themselves to a specific software stack.
Vidur~\citep{vidur} and Revati~\citep{revati}, for example, are tied to Sarathi-Serve~\citep{sarathi-serve} and FlashInfer~\citep{flashinfer}, while LLMServingSim~\citep{llmservingsim,llmservingsim2.0} is tied to HuggingFace or vLLM.
Each relies on a per-family model template that maps parameters such as \textit{head\_dim} and \textit{num\_heads} onto its predefined operator set.
Consequently, the simulators cover only a single column of the configuration matrix (\autoref{fig:configuration_space}), and any software stack update (\eg a changed module signature or a new fused operation) invalidates existing profiles.
Expanding coverage requires manual effort: \eg patching Vidur to support vLLM required understanding its parallelism initialization logic, integrating its custom fused kernels, and rewriting module callables to match Vidur's expected structure.
The unsustainability of this approach is reflected in GitHub issues for Vidur~\citep{vidur-issue-1, vidur-issue-2, vidur-issue-3}, LLMServingSim~\citep{llmserving-sim-issue-1}, and Apex~\citep{apex-issue-1}.

\mypara{High profiling overhead.}
Despite their restricted configuration scope, today's simulators incur substantial profiling overhead.
To construct an accurate latency model, each model requires an exhaustive sweep over a grid of batch sizes, sequence lengths, and context lengths up to the model's maximum context window, repeated for every new model or backend.

Consider a practitioner deploying a chat application and evaluating FlashInfer on vLLM across four sub-8B models: Llama-3.1-8B-Instruct, BioMistral-7B-Instruct, Qwen2-7B-Instruct, and DeepSeek-R1-Distill-Qwen-7B.
Vidur profiles 11 modules per model: attention modules are swept over sequence lengths, batch sizes, and context lengths, while non-attention modules are swept over sequence lengths alone.
Across the four models, this totals 44 module profiling runs and 27.7 wall-clock hours on a single A100 GPU.
While additional hardware reduces wall-clock time, the linear scaling with model count directly undermines the rapid iteration that simulation is meant to enable.
\section{Overview of \sys{}}
\label{sec:overview}

We propose \sys{}, a framework that explores the full configuration space while reducing profiling cost in LLM inference simulators by reusing profiles across configurations without manual setup.
We build on two structural observations: (1) every input dimension derives from either model configuration or the incoming request, and (2) model-configuration values recur across models, so the same operation runs in many configurations.
Rather than relying on manually defined templates or hardcoded operation sets, \sys{} \emph{dynamically extracts} the operations invoked for a given target model and configuration through a single inference pass with a dummy prompt, and \emph{selectively profiles} only operations not already in its latency database.

\subsection{Challenges}
\label{sec:challenges}
Realizing dynamic extraction and selective profiling in practice presents two non-trivial challenges.

\mypara{C1: Determining the semantic meaning of input parameters.}
Sweep-range selection and redundancy detection both require knowing what each tensor dimension represents, yet raw tensor shapes do not encode this.
Consider an \texttt{aten::matmul} with traced input shapes [60, 2048] and [2048, 4096]: 60 is a request-dependent batch size, while 2048 and 4096 are fixed by the model configuration.
Without this distinction, sweeping all dimensions explodes the profiling space, and dimension-level comparison across configurations fails: Qwen-3-8B and Llama-3-8B share identical attention geometry (32 query heads, 8 KV heads, 128 head dim), making their attention operations equivalent, yet tokenizer differences yield different token counts that flag them as distinct.

\mypara{C2: Isolating stateful operations.}
To profile per-operation latency, each operation must be runnable in isolation.
Existing tools such as the PyTorch Profiler extract the call graph of operations and their tensor inputs, but do not capture the runtime state that certain operations depend on.
Attention operations, for example, require attention metadata, KV cache memory addresses, and request-to-slot mappings, none of which are recorded in the trace.
Without this context, such operations cannot be replayed standalone even when all tensor inputs are available, and thus cannot be profiled without running the entire model.
Manually reconstructing this context demands deep knowledge of the serving engine's internals and is time-consuming and error-prone.

\subsection{Key Ideas}
To address these challenges, \sys{} introduces two key ideas.

\mypara{I1: Assigning semantic meaning to tensor dimensions via taint propagation.}
Since every input dimension derives from either model configuration or the incoming request, \sys{} taints values at the serving engine's model-configuration and request-processing entry points and propagates them through a single inference pass, labeling each dimension as model-derived (fixed across workloads) or request-derived (variable).
These labels enable principled sweep-range selection and correct identification of redundant operations across configurations.

\mypara{I2: Isolating context-dependent operations via hierarchical context resolution.}
Serving engines already perform the necessary context initialization during their own startup sequence; \sys{} reuses this code directly, reconstructing stateful operations in their proper execution environment without manual effort.
When an operation cannot execute in isolation, \sys{} falls back to its enclosing parent module in the PyTorch hierarchy, which already holds the initialized state.
This ensures every operation is profiled at the finest granularity possible: standalone when self-contained, or within its enclosing module when context is required.

\begin{wrapfigure}{r}{0.55\textwidth}
    \centering
    \vspace{-15px}
    \hspace{-10px}
    \includegraphics[width=\linewidth]{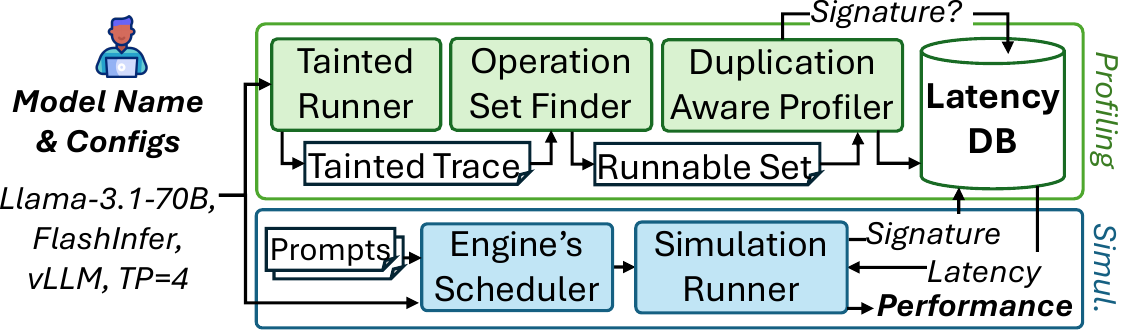}
    \caption{
        \sys{} architecture.
        }
        \label{fig:dooly_sys_design}
        \vspace{-12px}
    \end{wrapfigure}
\mypara{Workflow.}
\autoref{fig:dooly_sys_design} illustrates how \sys{} incorporates the two ideas.
The user provides a target model name and a set of configuration tuples to explore.
In the profiling phase, the Tainted Runner (\autoref{sec:tainted_runner}) performs a single inference pass on a dummy prompt and annotates the resulting PyTorch execution trace with taint labels for each tensor dimension, producing a \textit{tainted trace}.
The Operation Set Finder (\autoref{sec:op_set_finder}) parses this trace into a hierarchical call graph and resolves each operation to the finest profiling granularity.
The Duplication-Aware Profiler (\autoref{sec:duplication_aware_profiler}) then uses the semantic labels to determine per-operation sweep ranges and eliminate redundant operations across configurations, populating the latency database with only the necessary measurements.
The simulation phase (\autoref{sec:eval-accuracy}) reuses the serving engine's own scheduler to drive end-to-end simulation; for each scheduled request, the simulation runner walks the operation call graph and queries per-signature regression models trained on the latency database.
\section{Tainted Runner}
\label{sec:tainted_runner}
The Tainted Runner executes a single inference pass and produces a semantically annotated trace in which every tensor dimension is labeled with its origin.
It operates in two phases: \textbf{source tainting} intercepts the serving engine's entry points to initialize taint labels for model configuration and request parameters, and \textbf{taint propagation} tracks how these labels flow through arithmetic and tensor operations across the pass, an approach inspired by information flow tracking in prior systems~\cite{taintdroid,flowdroid,dta}.
\sys{} also emulates the parallelism setup and replaces collective operations with no-op hooks so that multi-GPU configurations can be traced on a single GPU.

\subsection{Source Tainting}
\label{sec:source_tainting}
Serving engines expose model- and request-specific parameters through well-defined dataclasses (\eg \texttt{SchedulerOutput} and \texttt{ModelConfig} in vLLM), whose member values \sys{} recursively \emph{taints}.
Model configuration values receive a \mconfig{} taint; request-dependent values receive \numtoks{} for token counts and \numreqs{} for the number of requests in the batch.
These source taints initialize the \textit{global taint registry}, which maps concrete values back to their taint labels.

\subsection{Taint Propagation}
\label{sec:taint_propagation}

\begin{wraptable}{r}{0.52\linewidth}
    \vspace{-12pt}
    \centering
    \small
    \caption
    {
        Scalar taint propagation rules.
        Taints ($\tau$) are either untainted ($\bot$), a base label ($l \in L$ where $L = \{\mconfig, \numtoks, \numreqs\}$), or composite ($\mix(H)$) where $H$ maps values to base taints.
        }
        \label{tab:taint-rules}
        \begin{tabular}{@{}lll@{}}
            \toprule
            \textbf{Operation} & \textbf{Result} & \textbf{Rule} \\
            \midrule
            $\bot \otimes \tau$ & $\tau$ & Absorption \\
            $\tau \otimes \tau$ & $\tau$ & Preservation \\
            $l_1 \otimes l_2$ {\scriptsize($l_1 \neq l_2$)} & $\mix(\{l_1, l_2\})$ & Conflict \\
            $\mix(H) \otimes l$ & $\mix(H \cup \{l\})$ & Extend \\
            $\mix(H_1) \otimes \mix(H_2)$ & $\mix(H_1 \cup H_2)$ & Merge \\
            \bottomrule
        \end{tabular}
        \vspace{-12pt}
\end{wraptable}

When arithmetic combines tainted scalars from different sources, the result's taint records all contributing origins (\eg \texttt{batch\_size} $\times$ \texttt{head\_dim} records both \numtoks{} and \mconfig{}).
\autoref{tab:taint-rules} formalizes the propagation rules, which extend naturally to tensor dimensions since shape computations are arithmetic over dimension sizes.
\sys{} categorizes tensor operations as \emph{dimension-mapping} or \emph{dimension-preserving}.

\mypara{Dimension-mapping operations} create new tensors or rearrange existing shapes (\eg \texttt{torch.empty}, \texttt{torch.reshape}).
These require explicit hooks because they directly establish the taint-to-dimension relationship.
For tensor creation (\eg \texttt{torch.empty}), \sys{} tags each output dimension with the taint of its corresponding scalar argument.
For tensor manipulation operations (\eg \texttt{torch.reshape}), the input parameters explicitly define the permutation and merge/split behavior of dimensions.
When dimensions merge, \sys{} applies the Conflict, Extend, or Merge rules to assign a \mix{} taint with $H$ recording the value-to-taint mapping.
When dimensions split, it recovers the original taints by consulting $H$.
Newly tainted dimensions are registered in the global taint registry, and \sys{} consults this registry to resolve inferred dimensions (\eg \texttt{reshape(-1)}).

\mypara{Dimension-preserving operations} maintain input shapes and thus preserve dimension taints.
Rather than instrumenting each operation individually, \sys{} intercepts these via PyTorch's Dispatcher.
This lets \sys{} observe operations \emph{after} dispatch but \emph{before} kernel execution, providing full access to input parameters and output tensor shapes.
This dispatch-level interception covers all operations without per-operation hooks, including custom kernels in attention backends and fused operations in serving engines.
For each intercepted operation, \sys{} applies shape-matching heuristics: output dimensions that match an input dimension inherit its taint, and any new dimensions are resolved via the global taint registry.\footnote{Taint ambiguity arises when multiple dimensions share the same concrete value but carry different taints. This is rare in practice and is resolved by retracing with a different prompt; see \autoref{sec:app-ambiguity} for details.}
\sys{} assigns a $\bot$ taint to unresolved dimensions, which can later be recovered via reshape operations if they merge with known dimensions.

\mypara{Trace annotation.}
Together, these rules tag every dimension with its source taint as the inference pass executes.
\sys{} records these taints via the PyTorch Profiler~\cite{pytorch_profiler, perfetto}, capturing them at multiple levels of the PyTorch~\cite{pytorch} stack: module hierarchy, high-level operations, and low-level ATen operators (see \autoref{sec:app-trace-annotation} for hook details).
This ensures all operations are semantically labeled within the call graph, including custom kernels that bypass the dispatcher.
We evaluate the effect of each rule in~\autoref{sec:taint-coverage}.
\section{Operation Set Finder}
\label{sec:op_set_finder}
The tainted trace from \autoref{sec:tainted_runner} contains every operation invoked during the model's forward pass, but not all of them can be profiled in isolation.
The Operation Set Finder resolves this by parsing the trace into a hierarchical call graph, testing each operation for standalone execution, and constructing the minimal runnable set covering all kernel calls in the pass.

\subsection{Hierarchy Constructor}
\label{sec:hierarchy_constructor}
\sys{} parses the tainted trace into a tree capturing the module-to-operation nesting of the model; the hierarchy is derived from the recorded durations of modules, operations, and kernels in the trace.
Each node corresponds to an \texttt{nn.Module} invocation (\eg \texttt{model.layers.0.self\_attn}), an ATen operator, or a CUDA kernel recorded within that module.
Because transformer architectures repeat the same layer structure, \sys{} prunes the tree across layers: structurally identical subtrees (\eg \texttt{layers.0.self\_attn} and \texttt{layers.1.self\_attn}) are collapsed into a single canonical subtree, reducing the resolution workload to one representative per repeated module.

\subsection{Testing for Runnability}
\label{sec:testing_runnability}
The hierarchy provides a fallback structure for resolution: if a leaf operation cannot run in isolation, \sys{} falls back to its parent module.
For any module to execute standalone, \sys{} must both reconstruct the necessary execution context and generate valid input tensors; it does so through execution context emulation and taint-driven input generation, respectively.

\mypara{Execution context emulation.}\label{sec:context-emulation}
Serving engines establish context in two stages.
At initialization, the engine binds module-level state such as KV cache memory and metadata builders for attention modules.
At each forward pass, it generates per-request context based on the batch's characteristics (sequence lengths, batch size).
This per-pass context is backend-dependent: different attention backends (\eg FlashInfer or FlashAttention) impose distinct memory layouts and computation strategies, requiring different metadata and execution contexts.
\sys{} faithfully emulates both stages by reusing the serving engine's internal APIs, injecting the constructed context (\eg via \texttt{set\_forward\_context()} in vLLM) to enable isolated module execution.

\mypara{Input generation.}
\sys{} uses the taint annotations to determine the shapes and values of input tensors.
For stateless modules (\eg linear layers, embeddings), tensors are generated directly from the tainted argument values in the trace.
For stateful modules (\eg Attention, Mamba, MoE layers), the same strategy applies, but context setup is delegated to the initialized engine.
In both cases, the following rules govern taint-driven input generation:
\mconfig{}-tainted dimensions remain fixed, as they define the model's architecture.
\numtoks{}- or \numreqs{}-tainted dimensions are set to small values to verify runnability with minimal overhead.
\mix{}-tainted dimensions are recalculated by substituting the workload-dependent component while preserving the model-derived component recorded in $H$; for example, a flattened dimension $\numtoks(269)\times\mconfig(40)=10760$ becomes $1 \times 40 = 40$.
Untainted dimensions (typically size-\texttt{1} vector dimensions) are left unchanged to preserve operator semantics.
The Duplication-Aware Profiler (\autoref{sec:duplication_aware_profiler}) reuses this input generation strategy when sweeping across workload configurations.

\mypara{Bottom-up resolution.}
With the hierarchy, context emulation, and input generation in place, \sys{} performs a bottom-up traversal to find the runnable set.
Resolution starts at tainted leaf nodes, skipping untainted leaves introduced by dispatch mechanics (\autoref{sec:taint_propagation}).
Starting at the finest granularity maximizes deduplication in the profiling stage (\autoref{sec:duplication_aware_profiler}), since fine-grained operations are more likely to be shared across models and configurations.

Each leaf is tested for standalone execution by (1) importing its operator(s) from the relevant libraries and (2) running them with the generated input tensors.
Leaves that pass both steps are added to the runnable set; those that fail trigger a fallback to the parent node.
Common failure cases are context-dependent operations (\eg \texttt{unified\_attention}), whose required runtime state cannot be recovered from the trace alone.
Stateless parents undergo the same import-and-run test as leaves, while stateful parents additionally require context emulation.
\sys{} identifies stateful modules by consulting the serving engine's registries (\eg vLLM's \texttt{AttentionGroup} registry).
If the parent cannot be resolved, the fallback continues upward until either a node succeeds or the tree root is reached.

When a parent resolves successfully, its children are removed from the runnable set to prevent double-counting.
This coarsening reduces deduplication opportunities, but it occurs only when isolated execution fails, since leaf-level resolution is always attempted first.
The resulting \emph{runnable set} is forwarded to the Duplication-Aware Profiler (\autoref{sec:duplication_aware_profiler}).

\subsection{Runnable Set Characteristics}
\label{sec:runnable_set}
The runnable set produced by this resolution process is heterogeneous in granularity.
Standalone-runnable operations (\eg \texttt{aten::linear}) appear as operator-level entries.
Operations whose isolated execution fails (typically context-dependent kernels or Triton implementations that cannot be imported standalone) are absorbed into their nearest enclosing module and appear as module-level entries carrying the necessary execution context.
This mixed-granularity construction lets \sys{} produce a complete runnable set for any architecture PyTorch can execute, including those whose dominant operators are not exposed as importable kernels.

\mypara{Sufficiency of a single trace.}
A single dummy-prompt trace suffices for full operation coverage across both prefill and decode phases.
Although the two phases invoke different code paths in context-dependent modules, the divergence is driven by attention metadata that \sys{}'s input generation injects per sweep point, exercising both paths at profile time; see \autoref{sec:operation-set-finder} for details.

\section{Duplication-Aware Profiler}
\label{sec:duplication_aware_profiler}
The runnable set from \autoref{sec:op_set_finder} defines the operations to profile, but profiling every operation for every configuration is wasteful: many operations are semantically equivalent across models and configurations.
The Duplication-Aware Profiler detects these equivalences before profiling.

\mypara{Signature generation.}
\sys{} formalizes equivalence through \textit{signatures}, where each signature identifies an operation by what is invariant across workloads.
A signature comprises three components: (1) the operation name and the values of its \mconfig{}-tainted dimensions and scalars, which are fixed by the model architecture; (2) the set of GPU kernel symbols dispatched during execution, which captures the backend's compile-time kernel selection; and (3) for module-level entries, a digest of the module instance's primitive Python attributes (\eg \texttt{sliding\_window}), which captures runtime branching invisible at the kernel level.
Workload-dependent dimensions (\numtoks{}, \numreqs{}) are excluded because they parameterize the sweep, not the operation's identity.
\sys{} computes the \emph{signature hash} as SHA-256 over a canonical serialization of these three components; the resulting hash serves as the operation's primary key in the latency database.
Two operations collapse to the same signature only when all three components match; \sys{} prunes the runnable set by skipping any entry whose signature hash is already present in the database from a prior session.

\mypara{Taint-driven sweep.}
For each remaining operation, \sys{} sweeps the \numtoks{}- and \numreqs{}-tainted dimensions over user-defined ranges of sequence lengths and batch sizes, keeping \mconfig{}-tainted dimensions fixed.
Input tensors are constructed using the taint-driven generation strategy from \autoref{sec:testing_runnability}, with \mix{}-tainted dimensions recalculated at each sweep point.
The resulting measurements are stored in the latency database, keyed by signature hash and workload configuration, providing the training data from which downstream simulators build per-signature regression models for latency prediction.

\mypara{Latency database.}
Profiling results are persisted in a SQLite database keyed by signature hash and workload configuration, making deduplication a primary-key lookup.
Communication operations live in a separate sub-schema indexed by hardware topology and tensor-parallel degree, since their latency does not depend on model architecture.
See \autoref{sec:app-database} for the full schema.

Together, the Tainted Runner, Operation Set Finder, and Duplication-Aware Profiler form \emph{\sys{}Prof}.
\section{Evaluation}
\label{sec:eval}

We evaluate \sys{} on simulation accuracy (\autoref{sec:eval-accuracy}), profiling overhead reduction (\autoref{sec:savings}), and taint coverage correctness (\autoref{sec:taint-coverage}).
We span the configuration space introduced in \autoref{sec:configurations}: model architectures across dense, GQA, MoE, and interleaved sliding-window attention families; three attention backends (FlashInfer, FlashAttention, TritonAttention) and two serving engines (vLLM and SGLang); real-trace (ShareGPT4~\cite{sharegpt4}) and synthetic prefill-heavy and decode-heavy workloads; and two hardware platforms (NVIDIA A100 and H100).

\mypara{\sys{}Sim implementation.}
To evaluate \sys{}, we build a simulator called \emph{\sys{}Sim} that predicts per-iteration inference latency from per-signature regression models trained on \sys{}'s latency database, using the same features as Vidur~\citep{vidur}.
The simulator comprises a scheduler that reuses the serving engine's logic to reproduce request batches, and a latency estimator that walks the model's call graph and queries the per-signature regression models.
We implement \sys{}Sim for vLLM; existing simulators~\citep{vidur,revati,llmservingsim,llmservingsim2.0,accelsim} can train per-signature regression models from \sys{}'s standalone latency database as a drop-in profiling backend.

\subsection{\sys{}Sim: Simulation Accuracy}
\label{sec:eval-accuracy}
\begin{figure}[h]
  \centering
  \vspace{-5px}
  \includegraphics[width=1.0\linewidth]{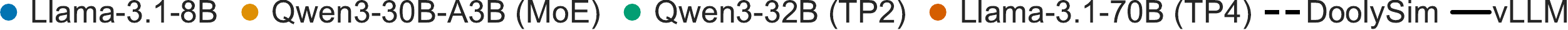}\\
  \vspace{-2pt}
  \begin{subfigure}[t]{0.33\linewidth}
    \centering
    \includegraphics[width=\linewidth]{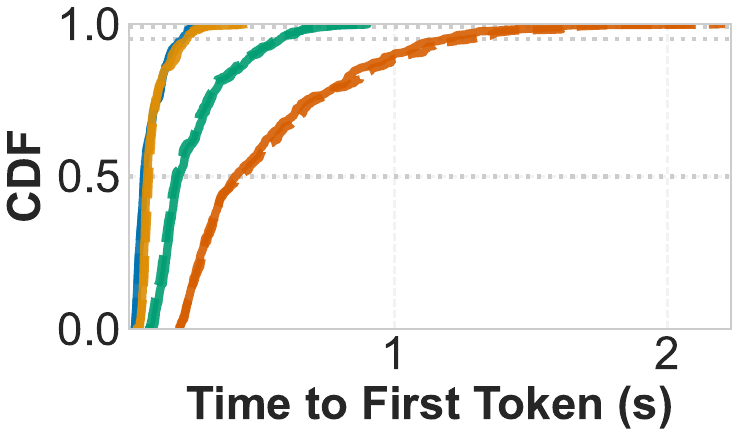}
    \caption{TTFT Predicted vs Measured}
    \label{fig:sim_accuracy:ttft}
  \end{subfigure}\hfill
  \begin{subfigure}[t]{0.33\linewidth}
    \centering
    \includegraphics[width=\linewidth]{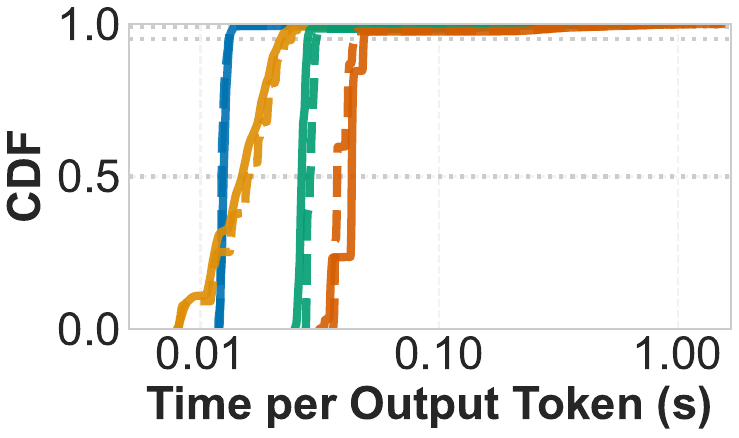}
    \caption{TPOT Predicted vs Measured}
    \label{fig:sim_accuracy:tpot}
  \end{subfigure}\hfill
  \begin{subfigure}[t]{0.33\linewidth}
    \centering
    \includegraphics[width=\linewidth]{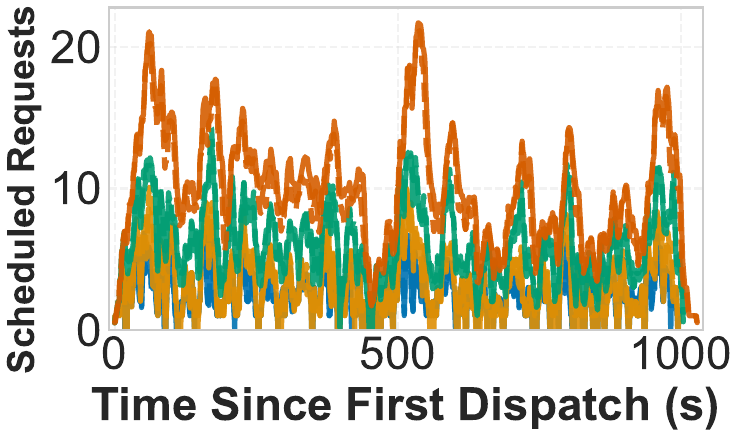}
    \caption{Scheduled requests over time}
    \label{fig:sim_accuracy:sched}
  \end{subfigure}
  \caption{
    \sys{}Sim accuracy on four model architectures with vLLM and FlashInfer on an A100 GPU, using ShareGPT4 at 0.5 requests per second.
    }
    \label{fig:sim_accuracy}
\end{figure}

\vspace{-10pt}
\mypara{End-to-end performance and scheduling.}
\sys{} predicts vLLM's measured latency with MAPE below 5\% for TTFT (\autoref{fig:sim_accuracy:ttft}) and below 8\% for TPOT (\autoref{fig:sim_accuracy:tpot}) across all reported percentiles.
The higher TPOT error reflects the small absolute magnitude of per-token decode latency: sub-millisecond computation and measurement jitter translate into disproportionately large relative error.
\autoref{fig:sim_accuracy:sched} shows that \sys{} faithfully emulates vLLM's batch scheduling decisions, yielding throughput and latency predictions with MAPE below 0.5\%.
\sys{}'s operation-level accuracy thus aggregates into reliable whole-system predictions.

\begin{figure}[h]
  \vspace{-5px}
  \centering
  \includegraphics[width=0.9\linewidth]{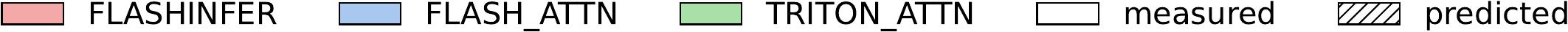}\\
  \begin{subfigure}[t]{0.49\linewidth}
    \centering
    \includegraphics[width=\linewidth]{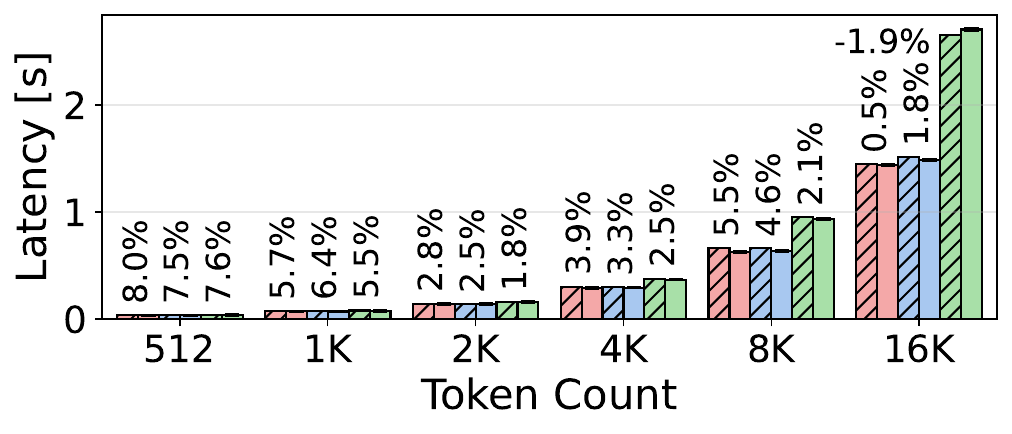}
    \vspace{-15px}
    \caption{A100}
    \label{fig:eval_config_space:a100}
  \end{subfigure}\hfill
  \begin{subfigure}[t]{0.49\linewidth}
    \centering
    \includegraphics[width=\linewidth]{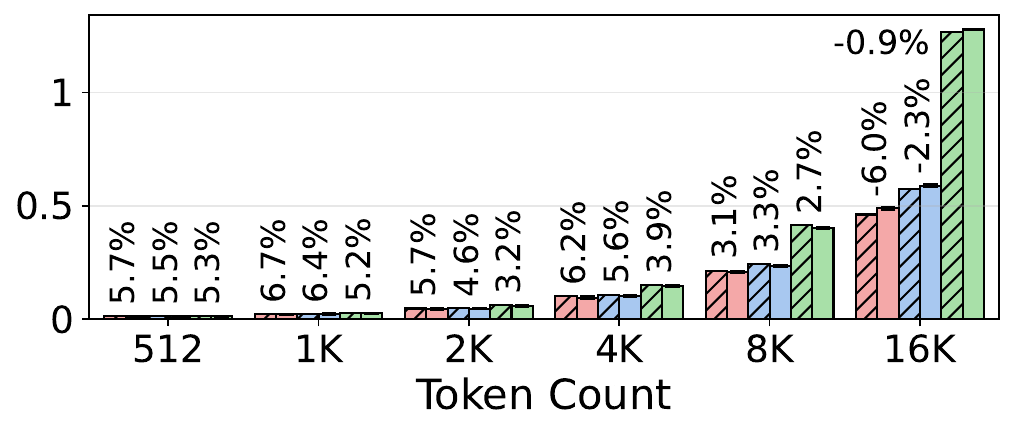}
    \vspace{-15px}
    \caption{H100}
    \label{fig:eval_config_space:h100}
  \end{subfigure}
  \vspace{-5pt}
  \caption{Per-batch latency predictions for three attention backends on A100 and H100, using Llama-3.1-8B.}
  \label{fig:eval_config_space}
\end{figure}

\vspace{-10pt}
\mypara{Configuration coverage.}
To validate that \sys{}Sim expresses the configuration space of \autoref{sec:configurations}, we simulate the prefill-heavy workload on Llama-3.1-8B across three attention backends on both A100 (\autoref{fig:eval_config_space:a100}) and H100 (\autoref{fig:eval_config_space:h100}) GPUs.
\sys{} predicts TTFT within 0.5--8.0\% error across all configurations, with the largest errors at small batch sizes, which are most sensitive to millisecond jitter.
This per-operator accuracy lets \sys{}Sim correctly reproduce the inversion points across model and backend selections on both hardware platforms and on the decode-heavy workload.
See \autoref{sec:app-sim-accuracy} for results on different models, hardware, and workloads, \autoref{sec:app-inversion} for inversion point analysis, and \autoref{sec:app-parallelism} for experiments with pipeline parallelism, expert parallelism, and others.

\subsection{\sys{}Prof: Profiling Overhead Analysis}
\label{sec:savings}
We measure profiling savings across a realistic corpus of 12 models in the 7B–8B parameter range, each profiled with three attention backends (FlashInfer, FlashAttention, TritonAttention), yielding 36 configurations.
The corpus spans three attention-architecture families: multi-head attention (Llama-2-7B, DeepSeek-LLM-7B-base), grouped-query attention (Llama-3/3.1-8B, Qwen2.5/3-8B, Mistral-7B-v0.3), and interleaved sliding-window attention (Ministral-8B-Instruct, Command-R7B).
All experiments required no modification to \sys{}Prof or \sys{}Sim to support new backends, models, or hardware, demonstrating out-of-the-box support across the configuration space.


\vspace{-10pt}
\begin{figure}[H]
  \centering
  \begin{minipage}[c]{0.42\linewidth}
    \centering
    \includegraphics[width=\linewidth]{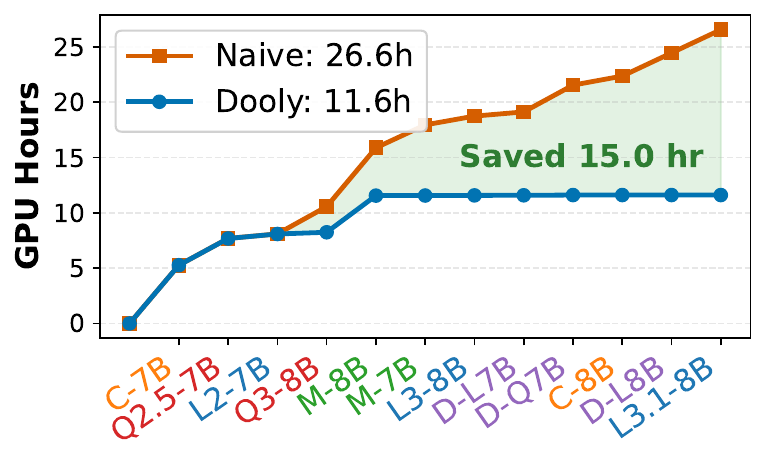}
    \vspace{-20pt}
    \captionof{figure}{
      Deduplication in \sys{} reduces profiling cost by 56\% by sharing latency measurements across configurations.
      Model labels are defined in~\autoref{sec:app-profiling-overhead}.
    }
    \label{fig:overlap}
  \end{minipage}
  \hspace{0.03\linewidth}%
  \begin{minipage}[c]{0.54\linewidth}
    \centering
    \vspace{15pt}
    \vspace{-10pt}
    \footnotesize
    \setlength{\tabcolsep}{1.8pt}
    \renewcommand{\arraystretch}{0.9}
    \begin{tabular}{@{}llrrrrr@{}}
      \toprule
      \textbf{Group} & \textbf{Variant} & \textbf{N} & \textbf{R} & \textbf{Profile (h)} & \textbf{Saved (h)} & \textbf{Red. (\%)} \\
      \midrule
      \multirow{6}{*}{\textbf{Attention}} & \textit{(aggregate)} & 42 & 27 & 10.86 & 14.24 & 56.7 \\
      & 32/8/128 & 24 & 21 & 2.83 & 11.62 & 80.4 \\
      & 28/4/128 & 6 & 3 & 2.24 & 2.24 & 50.0 \\
      & 32/32/128 & 6 & 3 & 0.35 & 0.38 & 52.2 \\
      & window=4K & 3 & 0 & 2.24 & 0.00 & 0.0 \\
      & window=32K & 3 & 0 & 3.20 & 0.00 & 0.0 \\
      \midrule
      \multicolumn{2}{@{}l}{Linear (aten::linear)} & 180 & 153 & 0.54 & 0.56 & 50.8\\
      \multicolumn{2}{@{}l}{Other operators} & 390 & 332 & 0.18 & 0.19 & 50.8\\
      \midrule
      \multicolumn{2}{@{}l}{\textbf{Total}} & \textbf{612} & \textbf{512} & \textbf{11.58} & \textbf{14.98} & \textbf{56.4} \\
      \bottomrule
      \end{tabular}
    \captionsetup{type=table}
    \captionof{table}{
      Per-operation profiling savings.
      \textbf{N}: number of occurrences; \textbf{R}: number of reuses; \textbf{Profile}: total profiling latency; \textbf{Saved}: total hours saved; \textbf{Red.}: \% reduction.
    }
    \label{tab:profiling-savings}
    \vspace{5pt}
  \end{minipage}
\end{figure}
\vspace{-20pt}

As shown in \autoref{fig:overlap}, a naive per-configuration profiler would profile 612 operator signatures across the 36 configurations, taking 26.6 GPU-hours in total.
\sys{} identifies 100 unique signatures and deduplicates the rest automatically via call-graph-aware signature matching.
Profiling just 4 of the 12 models covers over 98\% of the total profiling time, with each of the remaining 8 models adding under 0.2 GPU-hours of new work.
This saturation is driven by attention, which dominates per-model profiling cost: \autoref{tab:profiling-savings} shows that 95\% of the 15 GPU-hours saved come from attention-signature reuse across architecturally similar models, while the remaining 5\% is distributed across other operations.
Once the first few models exercise the distinct attention configurations in the corpus, later models inherit their attention kernels and add negligible overhead.
\autoref{sec:app-profiling-overhead-extended} shows that this observation holds for larger models and MoE architectures as well.

\begin{wraptable}{r}{0.4\linewidth}
    \vspace{-10pt}
    \small
    \caption
    {
      Profiling cost (GPU-hours) of different methods across the same 12-model corpus.
      Savings are relative to naive profiling.
    }
    \label{tab:profiling-method-comparison}
    \begin{tabular}{@{}lrr@{}}
      \toprule
      \textbf{Method} & \textbf{Profile (h)} & \textbf{Saved (\%)} \\
      \midrule
      Naive & 26.54 & 0.0 \\
      HF-Diff & 24.44 & 7.9 \\
      HF-Model-Diff & 15.97 & 39.8 \\
      \sys{} & 11.58 & 56.4 \\
      \bottomrule
    \end{tabular}
    \vspace{-12pt}
\end{wraptable}

\mypara{Comparison with baselines.}
We compare \sys{} with three profiling strategies.
Naive profiling measures every configuration from scratch.
HF-Diff (HD) reuses a profile only when the full HuggingFace~\cite{huggingface} model configuration files match.
HF-Model-Diff (HMD) compares architectural parameters (\eg head count, layer count, hidden dimension) to identify reusable profiles across models with different configurations.
Across the original 36 configurations, \sys{} reduces profiling hours by 56.4\%, while HD and HMD reduce them by 7.9\% and 39.8\%, respectively.

HD and HMD fail to capture the models' operation-level structure that \sys{} exploits.
Consider two models with similar architectures but different attention types: Command-R7B (interleaved, sliding-window attention) and Aya-Expanse-8B (full attention).
HMD treats Command-R7B and Aya-Expanse-8B as equivalent because their compared architecture fields match.
Reusing the latter's profile for the former causes a 37.3\% error in TTFT prediction, while \sys{}'s operation-level tainting correctly identifies the two models' attention operations as distinct, yielding a 4.1\% error (see~\autoref{sec:app-baseline-accuracy} for more details).

\subsection{Taint Propagation: Coverage Analysis}
\label{sec:taint-coverage}
To validate that taint propagation correctly classifies each tensor dimension by source, we trace four models spanning diverse architectures: Llama-3.1-8B (dense), Command-R7B (heterogeneous), Qwen3-30B-A3B (MoE), and Llama-3.1-70B (tensor parallelism).
We hold the dummy prompt length constant and vary the batch size to inspect all \mconfig{}-, \numtoks{}-, and \numreqs{}-tagged dimensions in the taint registry, covering vLLM and SGLang, each paired with FlashInfer, FlashAttention, and TritonAttention.

Across all models, \mconfig{} values are constant across batch sizes while \numtoks{} and \numreqs{} scale exactly with the workload.
To stress-test the system, we deliberately use prompt lengths and batch sizes whose values collide with \mconfig{} values (\eg a batch size of 8 collides with the KV head count dimension); \sys{} detects these via conflicting taint assignments and resolves them by retracing with collision-free prompts.
Taint classification thus achieves 100\% accuracy in distinguishing model-derived from workload-derived dimensions.

\begin{wraptable}{r}{0.4\linewidth}
    \centering
    \small
    \caption
    {
      Ablation study of \sys{}'s taint-propagation rules for the two types of operations discussed in \autoref{sec:taint_propagation}.
      ``Forced'' suppresses the ambiguity failure.
    }
    \label{tab:taint-ablation}
    \setlength{\tabcolsep}{3pt}
    \renewcommand{\arraystretch}{1.1}
    \begin{tabular}{@{}lrr@{}}
    \toprule
    Trace variant & Untainted operations \\
    \midrule
    Full \sys{} & 0 \\
    w/o DP & (ambiguity failure) \ding{56}\\
    w/o DP, Forced & 241 \\
    w/o DM & 448 \\
    \bottomrule
    \end{tabular}%
    \vspace{-12pt}
\end{wraptable}

\mypara{Effect of taint propagation rules.}
The tainted runner relies on two propagation rules, one for each type of tensor operation: (1) dimension-mapping (DM) and (2) dimension-preserving (DP).
To measure their effect, we ablate each rule and inspect the resulting trace.
As shown in \autoref{tab:taint-ablation}, removing the dimension-mapping rule leaves 448 untainted operations, while removing the dimension-preserving rule causes a failure due to ambiguity in taint assignment.
When the ambiguity failure is suppressed, 241 operations remain untainted, demonstrating that both rules are essential for complete taint coverage.
These missing labels affect downstream profiling: an untainted request dimension may remain fixed instead of being swept, while an untainted model dimension is omitted from the signature used for deduplication.

\subsection{Operation Set Finder: Effect of Profiling Granularity}
\label{sec:profiling-granularity}
\sys{} produces a runnable set with mixed granularity: operations that run standalone are profiled at the \emph{operation level}, while context-dependent operations are profiled at the \emph{module level} (\autoref{sec:runnable_set}).
We compare three module--operation pairs.
The first pair is vLLM's \texttt{RowParallelLinear} module and the \texttt{aten::linear} operation it dispatches through the PyTorch Dispatcher; both execute the same cuBLAS GEMM.
The other two pairs are Gemma's RMSNorm and Cohere's LayerNorm modules and their \texttt{forward\_impl} operations, which \texttt{torch.compile} fuses into Triton kernels that bypass the Dispatcher.
Across all swept token counts, module- and operation-level latencies differ by less than 1\% for the linear pair and less than 0.1\% for both normalization pairs, within measurement jitter (\autoref{sec:app-granularity}).
Mixed-granularity profiling is therefore accurate for end-to-end simulation; the cost of falling back to a module is a lost deduplication opportunity, not a loss of accuracy.


\section{Applications of \sys{}}
\label{sec:applications}

Beyond the profile-based simulation use case, \sys{}'s configuration-agnostic profiling can be applied to other scenarios where operation-level profiling is needed across multiple configurations.
\mbox{llm-d}~\citep{llm-d, llm-d-heter}, a Kubernetes-native distributed inference framework, deploys model servers as pods and routes each request to the pod with the lowest predicted latency, a task that grows harder as the pool comes to span different models, GPUs, and serving configurations.
\sys{} complements such \emph{latency-prediction-based schedulers}: because its operation signatures are shared across configurations, measurements profiled once can be reused across pods, giving accurate TTFT and TPOT predictions for each one without profiling every pod separately.
Model engineers can also use \sys{} to \emph{benchmark new configurations} that share operations with previously profiled ones, speeding up model development by evaluating them without profiling from scratch.
\section{Limitations}
\label{sec:limitations}
Operations that bypass PyTorch's dispatcher (\eg ones implemented in Triton~\cite{triton}) can lack kernel-level taints; their enclosing modules, however, receive tainted inputs directly from the trace source, so the call graph remains fully covered.
\sys{}'s Tainted Runner currently traces text prompts, but taint propagation is modality-agnostic and extends to vision or audio by tainting their input sources; adapting it requires only identifying the new modality's request entry points.
Finally, \sys{} needs updates as serving engines evolve, but only source tainting is engine-specific (\eg vLLM v0.25.1 needed a 2-line fix), which is easier to maintain than a full engine-specific profiler.
\section{Conclusion}
\label{sec:concl}

We presented \sys{}, a configuration-agnostic, redundancy-aware profiling framework for LLM inference simulators.
\sys{} exploits the fact that operations share common model-specific parameters across configurations: taint propagation labels each tensor dimension at its source to enable cross-configuration deduplication, while hierarchical context resolution reuses the serving engine's logic to isolate stateful operations without manual instrumentation.
\sys{} lets practitioners evaluate serving strategies across a wide range of configurations at reduced profiling cost.
\textbf{Broader impacts:} We foresee no significant ethical or societal impact from this work.

\begin{ack}
We thank the anonymous reviewers for their constructive feedback, which helped improve this paper.
This work was supported by the U.S. National Science Foundation (NSF) under Award 2403026.
\end{ack}
\bibliographystyle{plainnat}
\bibliography{references}

\clearpage
\appendix

\section{Feature Comparison with Existing LLM Simulators}
\label{sec:app-checklist}
\definecolor{doolyhl}{RGB}{230,240,255}
\definecolor{yescolor}{RGB}{34,139,34}
\newcommand{\yes}{\textcolor{yescolor}{\ding{51}}}
\newcommand{\no}{\textcolor{nocolor}{\ding{55}}}
\newcommand{\maybe}{\textcolor{orange}{\ding{63}}}
\definecolor{nocolor}{RGB}{200,200,200}

\begin{table}[H]
\centering
\small
\setlength{\tabcolsep}{1.5pt}
\renewcommand{\arraystretch}{1.05}
\begin{tabular}{@{}l|>{\columncolor{doolyhl}}c|cccccc@{}}
\toprule
 & \textbf{\sys{}} & RT~\cite{revati} & VD~\cite{vidur} & FT~\cite{frontier} & LS\cite{llmservingsim} & AP~\cite{apex} & LS2~\cite{llmservingsim2.0} \\
\midrule
\rowcolor{gray!8}\multicolumn{8}{@{}l}{\textbf{\textit{Profiler}}} \\
\quad Automatic Model Discovery       & \yes & \no & \no & \no & \no & \no & \no \\
\quad Operator Set Extraction & \yes & \no & \no & \no & \no & \no & \no \\
\quad Automatic Profiling             & \yes & \no & \no & \no & \no & \no & \no \\
\addlinespace[2pt]
\rowcolor{gray!8}\multicolumn{8}{@{}l}{\textbf{\textit{Simulation}}} \\
\quad Continuous Batching & \yes & \yes & \yes & \yes & \yes & \yes & \yes \\
\quad Chunked Prefill     & \yes & \yes & \yes & \yes & \yes & \yes & \yes \\
\quad Prefix Caching      & \yes & \yes & \no  & \no  & \no  & \no  & \yes \\
\quad Heterogeneous Models  & \yes & \no & \no  & \no & \no  & \no & \no \\
\quad Mixture of Experts  & \yes & \yes & \no  & \yes & \no  & \yes & \yes \\
\bottomrule
\end{tabular}
\caption{
    Out-of-the-box capability comparison across profile-based LLM simulators.
    \sys{} (highlighted) automates model architecture discovery and profiling, while existing simulators require manual configuration to support new models or configurations (RT: Revati, VD: Vidur, FT: Frontier, LS: LLMServingSim, AP: Apex, LS2: LLMServingSim 2.0).
    }
\label{tab:checklist}
\end{table}

\vspace{-20pt}
\sys{}Prof automatically discovers the model architecture and extracts the operation set from a single trace, eliminating manual configuration when profiling new models or hardware.
By reusing the serving engine's scheduling logic, \sys{}Sim supports continuous batching, chunked prefill, and prefix caching out of the box.
Existing simulators, in contrast, require extensive manual configuration to profile new models or support new scheduling features, posing a significant barrier to adoption and maintenance as the ecosystem evolves.

\section{Resolving Taint Ambiguity}
\label{sec:app-ambiguity}
Taint ambiguity arises when multiple dimensions share the same concrete value but carry different taints; such collisions are rare in practice.
\mconfig{} values come from the model architecture and are typically powers of two (\eg 32, 8192); collisions \emph{within} \mconfig{} are benign because the colliding dimensions share the same taint.
\numtoks{} and \numreqs{} come from the dummy prompt, which \sys{} controls and sets to values unlikely to match any model-configuration dimension.
When a collision does occur, \sys{} detects it via conflicting taint assignments and retraces with a different prompt to resolve the ambiguity.
  
\section{Trace Annotation Hooks}
\label{sec:app-trace-annotation}
\sys{} augments the trace with taint information at three levels of the PyTorch stack:
\texttt{Module.\_\_call\_\_} records the enclosing module name (\eg \texttt{layers.0.self\_attn}), establishing the module hierarchy context for each operation;
\texttt{\_\_torch\_function\_\_} captures high-level operations with their original shape arguments before operator decomposition;
and \texttt{\_\_torch\_dispatch\_\_} intercepts the resulting low-level ATen operators.
Some operations, such as custom attention kernels, may execute outside the \texttt{\_\_torch\_dispatch\_\_} path and thus lack taint context at the kernel level; \sys{} preserves their semantic meaning at higher levels in the call graph, ensuring comprehensive labeling for downstream profiling.

\section{Bottom-Up Resolution and Single-Trace Sufficiency}
\label{sec:operation-set-finder}

\begin{wrapfigure}{r}{0.4\linewidth}
    \centering
    \includegraphics[width=\linewidth]{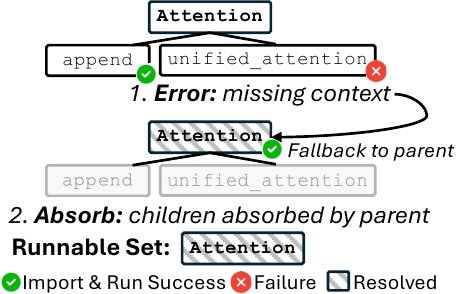}
    \caption{Bottom-up resolution process.}
    \label{fig:resolution}
\end{wrapfigure}

After extracting the operation set with the Tainted Runner, \sys{}Prof iterates through the call graph bottom-up, from leaf kernels to root modules (\autoref{fig:resolution}).
At each level, \sys{}Prof performs a dummy run with the tainted parameters to determine whether the operation can run in isolation or is context-dependent and requires module-level resolution.
When a parent becomes an anchor because its children are context-dependent, \sys{}Prof absorbs the children to avoid double-counting.
The dummy runs reuse the serving engine's initialization and forward contexts to set up the necessary metadata.

\vspace{10pt}

\mypara{Sufficiency of a single trace.}
A natural concern is whether a single dummy-prompt trace can cover both prefill and decode call paths.
Phase-dependent branching is driven entirely by token-count fields in the attention metadata that the serving engine passes through the forward context, and only context-dependent modules consume this metadata; all other operations in the runnable set are phase-invariant by construction.
\sys{} prepares context-dependent operations in their module form (\autoref{sec:context-emulation}), capturing both branches regardless of which phase the trace exercised, and its input generation mechanism injects the appropriate metadata for each sweep point at profile time.
A single trace therefore suffices for full operation coverage.

\section{Latency Database Schema}
\label{sec:app-database}
\begin{figure}[H]
    \centering
    \includegraphics[width=1.0\linewidth]{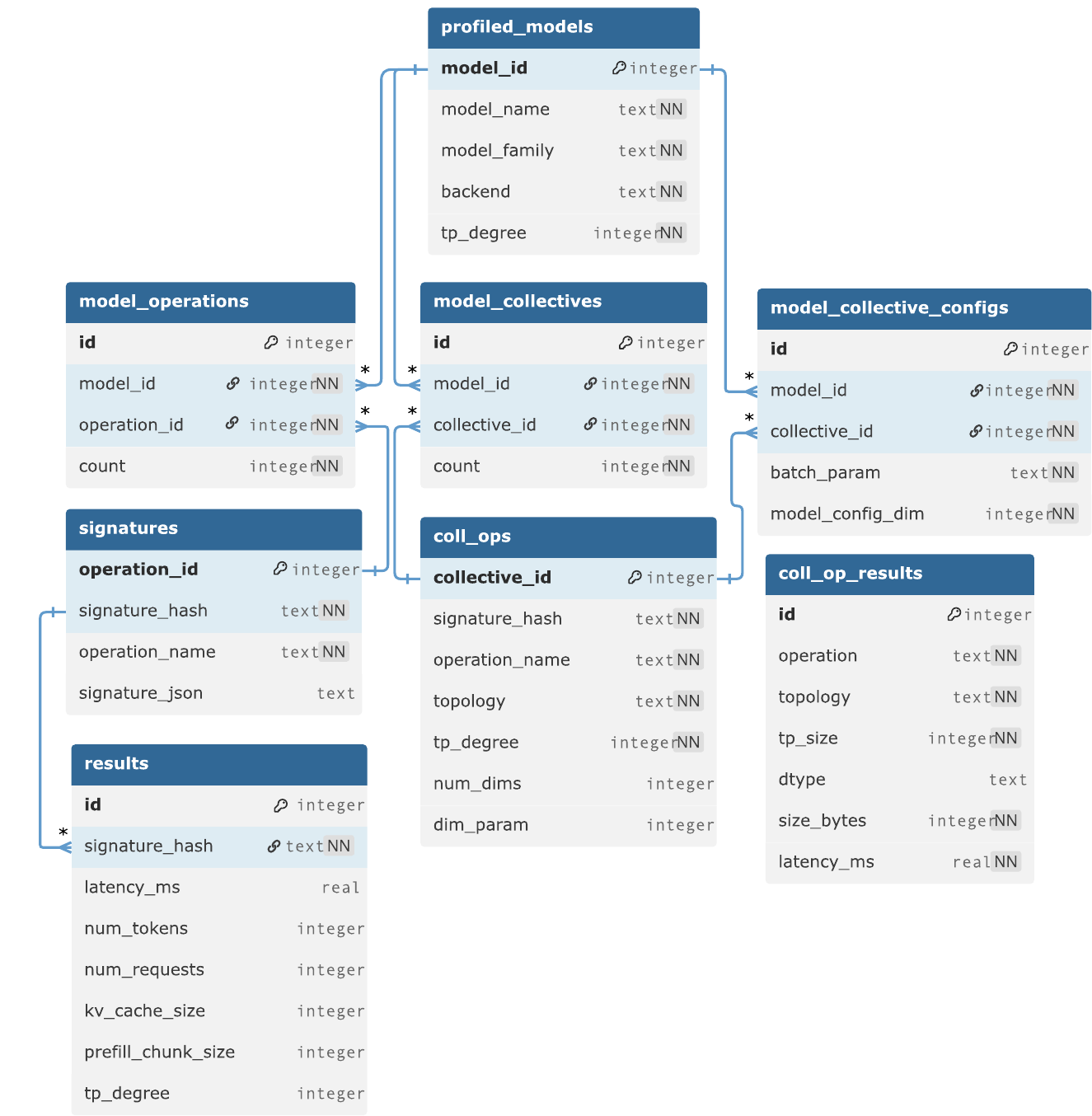}
    \caption{Latency database schema.}
    \label{fig:db-schema}
\end{figure}

\sys{} persists profiling results in a SQLite database organized around three orthogonal axes: profiled configurations (hardware $\times$ model $\times$ backend $\times$ tensor-parallel degree), unique operation signatures, and workload-dependent measurements.
The \texttt{model\_operations} join table maps each profiled configuration to the signatures it invokes, making deduplication a primary-key lookup against the \texttt{signatures} table.
Communication operations reside in a separate sub-schema because their latency depends on hardware topology and tensor-parallel degree rather than model architecture; they can therefore be reused across all configurations sharing the same \texttt{(topology, tp\_degree)}.
\autoref{fig:db-schema} shows the full schema.

\newpage
\newpage

\section{\sys{}Sim Simulation Accuracy}
\label{sec:app-sim-accuracy}

\subsection{End-to-End Setup and H100 Results}
\label{sec:app-e2e-accuracy}

\begin{table}[h]
  \centering
  \small
  \setlength{\tabcolsep}{6pt}
  \renewcommand{\arraystretch}{1.15}
  \begin{tabular}{ll}
    \toprule
    \multicolumn{2}{l}{\textbf{System configuration}} \\
    \quad Hardware            & 4$\times$ NVIDIA A100 GPU \\
    \quad Serving engine      & vLLM v0.17.1 (asynchronous scheduling) \\
    \quad Attention backend   & FlashInfer \\
    \quad Prefill chunk size  & 8{,}192 tokens \\
    \midrule
    \multicolumn{2}{l}{\textbf{Workload trace} ($>$8B parameters)} \\
    \quad Request rate        & 0.5\,req/s \\
    \quad Prompt length       & median $950$ / mean $1{,}232$ tokens \\
    \quad Output length       & median $388$ / mean $397$ tokens \\
    \midrule
    \multicolumn{2}{l}{\textbf{Models evaluated}} \\
    \quad Llama-3.1-8B            & Dense, 0.5\,rps trace \\
    \quad Qwen3-32B (TP=2)        & Dense, 0.5\,rps trace \\
    \quad Llama-3.1-70B (TP=4)    & Dense, 0.5\,rps trace \\
    \quad Qwen3-30B-A3B           & MoE, 0.5\,rps trace \\
    \bottomrule
  \end{tabular}
  \vspace{5pt}
  \caption{
    Experiment setup for evaluating the accuracy of \sys{}Sim on an A100 GPU.}
  \label{tab:exp-stream-setup-a100}
  \end{table}

\vspace{-15pt}

\begin{table}[h]
\centering
\small
\setlength{\tabcolsep}{6pt}
\renewcommand{\arraystretch}{1.15}
\begin{tabular}{ll}
\toprule
\multicolumn{2}{l}{\textbf{System configuration}} \\
\quad Hardware            & 1$\times$ NVIDIA H100 GPU \\
\quad Serving engine      & vLLM v0.17.1 (asynchronous scheduling) \\
\quad Attention backend   & FlashInfer \\
\quad Prefill chunk size  & 8{,}192 tokens \\
\midrule
\multicolumn{2}{l}{\textbf{Workload trace} ($\leq$14B parameters)} \\
\quad Request rate        & 1.0\,req/s \\
\quad Prompt length       & median $1{,}717$ / mean $1{,}911$ tokens \\
\quad Output length       & median $385$ / mean $410$ tokens \\
\midrule
\multicolumn{2}{l}{\textbf{Workload trace} ($>$14B parameters)} \\
\quad Request rate        & 0.5\,req/s \\
\quad Prompt length       & median $950$ / mean $1{,}232$ tokens \\
\quad Output length       & median $388$ / mean $397$ tokens \\
\midrule
\multicolumn{2}{l}{\textbf{Models evaluated}} \\
\quad Llama-3.1-8B            & Dense, 1.0\,rps trace \\
\quad Ministral-8B            & Heterogeneous, 1.0\,rps trace \\
\quad Qwen3-14B               & Dense, 1.0\,rps trace \\
\quad Mixtral-8$\times$7B     & MoE, 1.0\,rps trace, FP8 Quantized \\
\quad Qwen3-32B (TP=1)        & Dense, 0.5\,rps trace \\
\bottomrule
\end{tabular}
\vspace{5pt}
\caption{
  Experiment setup for evaluating the accuracy of \sys{}Sim on an H100 GPU.}
\label{tab:exp-stream-setup-h100}
\end{table}

\vspace{-10pt}
\mypara{Experiment setup.}
\autoref{tab:exp-stream-setup-a100} and \autoref{tab:exp-stream-setup-h100} summarize the experimental setups for the A100 and H100 evaluations, which share the same serving engine and scheduler configuration.
We drive the simulation with traces from the ShareGPT4 dataset.
Following Vidur~\cite{vidur} and Revati~\cite{revati}, we set the workload intensity just below each platform's maximum serving capacity: too low a rate under-utilizes the system and yields negligible scheduling delay, making latency predictions less meaningful; too high a rate overloads the system and rapidly inflates scheduling delay.

For latency estimation, \sys{}Sim trains one regression model per signature in the latency database.
The input features follow Vidur~\cite{vidur} and Revati~\cite{revati}: token count for non-Attention operations, and prefill token count, batch size, prefill chunk size, and pre-computed KV-cache size for Attention operations.
The pipeline is modular, so users can add new features for both profiling and estimation.

\begin{figure}[H]
  \centering
  \includegraphics[width=1.0\linewidth]{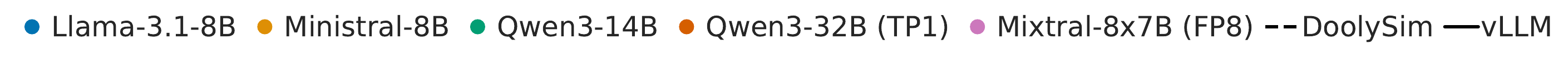}\\
  \vspace{-2pt}
  \begin{subfigure}[t]{0.33\linewidth}
    \centering
    \includegraphics[width=\linewidth]{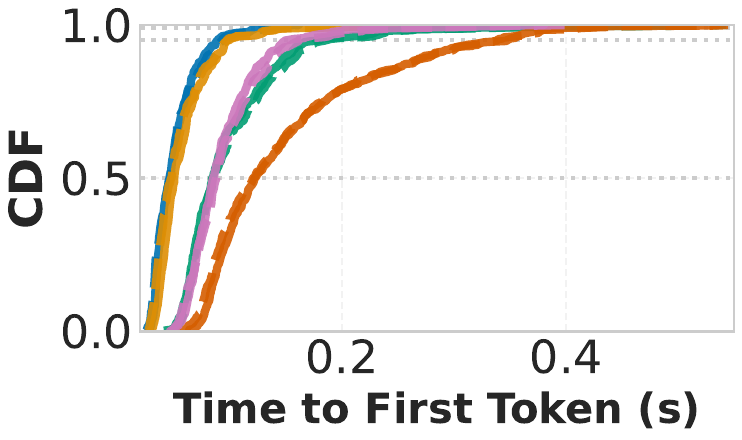}
    \caption{TTFT Predicted vs Measured}
    \label{fig:sim_accuracy:ttft-h100}
  \end{subfigure}\hfill
  \begin{subfigure}[t]{0.33\linewidth}
    \centering
    \includegraphics[width=\linewidth]{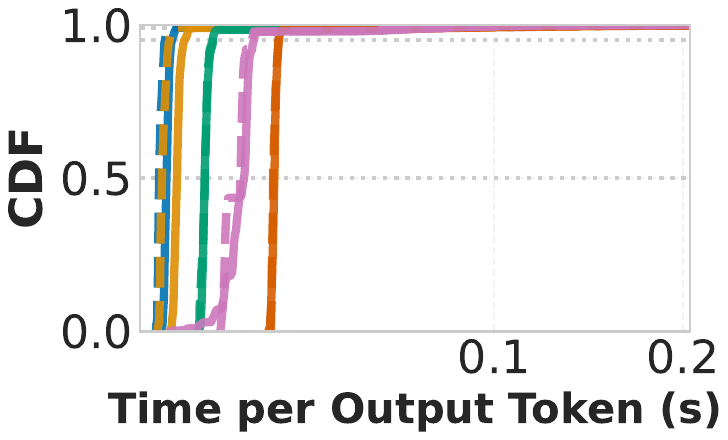}
    \caption{TPOT Predicted vs Measured}
    \label{fig:sim_accuracy:tpot-h100}
  \end{subfigure}\hfill
  \begin{subfigure}[t]{0.33\linewidth}
    \centering
    \includegraphics[width=\linewidth]{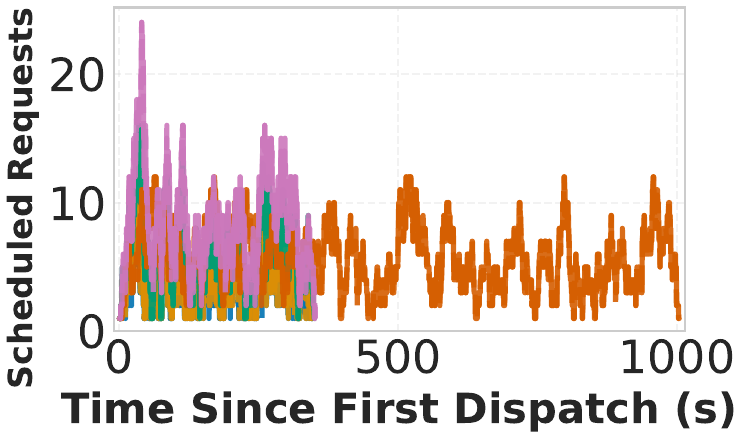}
    \caption{Scheduled requests over time}
    \label{fig:sim_accuracy:sched-h100}
  \end{subfigure}
  \caption{
    \sys{}Sim accuracy on five model architectures with vLLM and FlashInfer on an H100 GPU, using ShareGPT4 at 1.0 requests per second.
    }
    \label{fig:sim_accuracy_h100}
    \vspace{-10px}
  \end{figure}

\mypara{Results on an H100 GPU.}
We repeat the evaluation of \autoref{sec:eval-accuracy} on an H100 GPU (\autoref{fig:sim_accuracy_h100}).
The MAPEs of TTFT and TPOT predictions are 2\% and 5\%, respectively.
The simulated schedule closely matches the measured schedule, demonstrating \sys{}Sim's high fidelity in modeling latency and scheduling behavior on H100 GPUs as well.
\vspace{2pt}

\subsection{Per-Batch Accuracy Across Models, Hardware, and Workloads}
\label{sec:app-per-batch}

\mypara{Experiment setup.}
Instead of streaming requests, we run a single batch with a fixed size and measure its latency.
This isolates per-batch latency prediction accuracy from scheduling effects; per-batch accuracy is critical for accurately modeling the serving engine's scheduling behavior.
To prevent vLLM's scheduler from splitting the batch, we set the prefill chunk size larger than the batch's total token count.

\begin{figure}[H]
  \centering
  \vspace{-8px}
  \includegraphics[width=0.9\linewidth]{per_batch_legend.pdf}\\
  \begin{subfigure}[t]{0.49\linewidth}
    \centering
    \includegraphics[width=\linewidth]{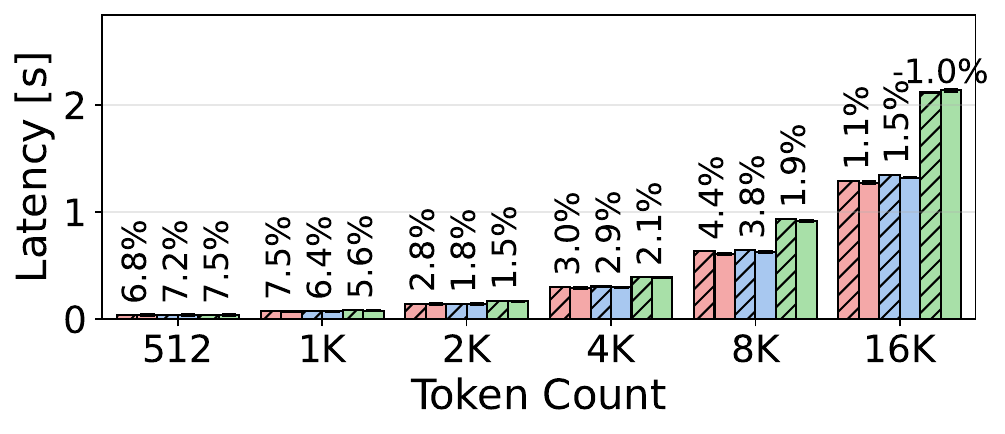}
    \caption{A100}
  \end{subfigure}
  \begin{subfigure}[t]{0.49\linewidth}
    \centering
    \includegraphics[width=\linewidth]{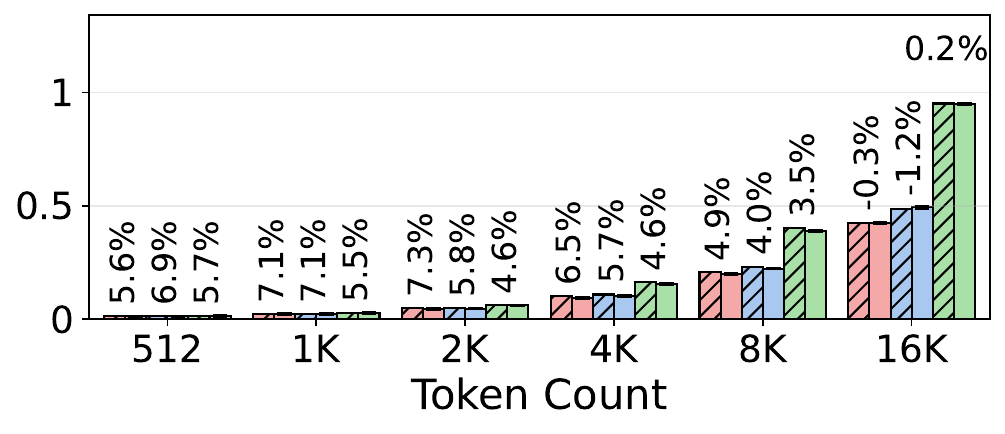}
    \caption{H100}
  \end{subfigure}
  \caption{
    \sys{}Sim accurately predicts the per-batch latency on Command-R7B with three attention backends on both A100 and H100 GPUs for a prefill-heavy workload.
    }
    \label{fig:per-batch-accuracy-prefill}
\end{figure}

\begin{figure}[H]
  \centering
  \vspace{-10pt}
  \includegraphics[width=0.9\linewidth]{per_batch_legend.pdf}\\
  \begin{subfigure}[t]{1.0\linewidth}
    \centering
    \includegraphics[width=\linewidth]{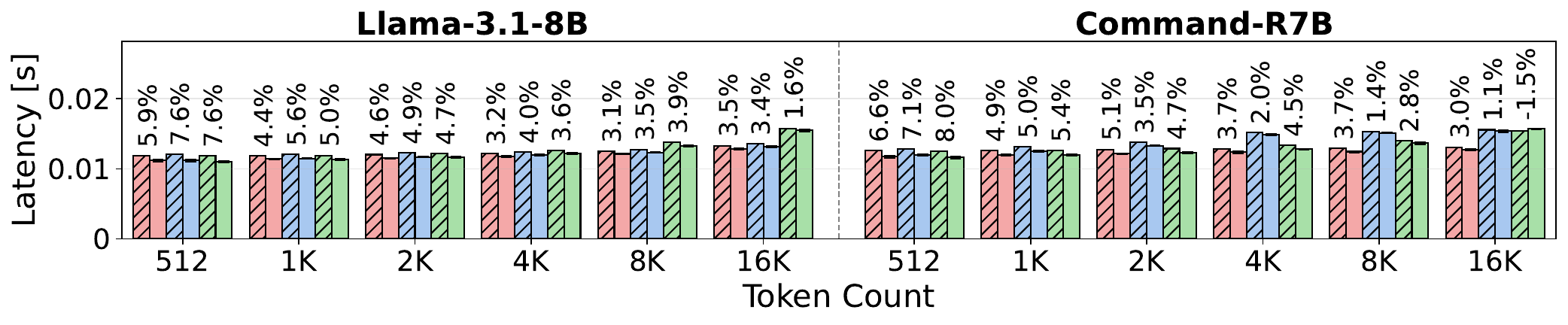}
    \caption{Decode-heavy workload on A100 GPU}
  \end{subfigure}
  \begin{subfigure}[t]{1.0\linewidth}
    \centering
    \includegraphics[width=\linewidth]{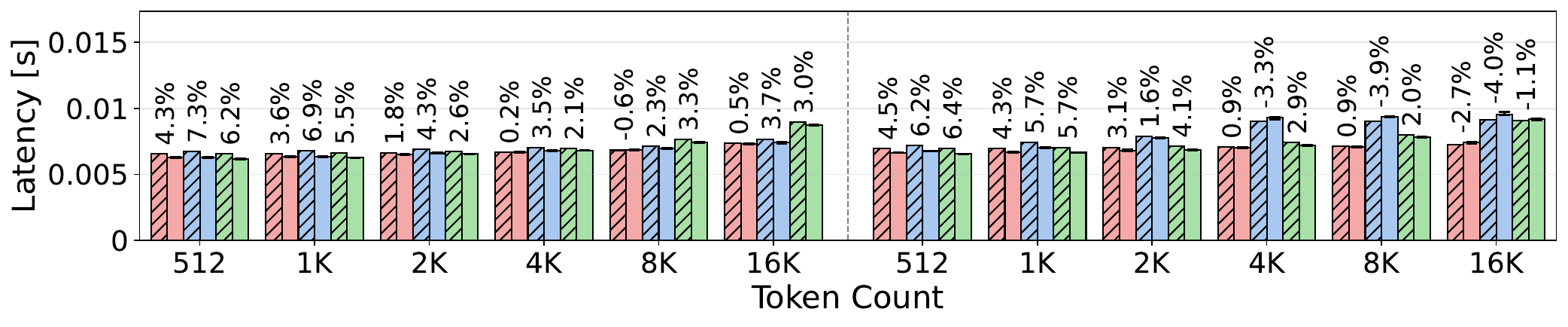}
    \caption{Decode-heavy workload on H100 GPU}
  \end{subfigure}
  \caption{
    \sys{}Sim accurately predicts the per-batch latency across different model and attention backend selections on A100 and H100 GPUs for a decode-heavy workload.
    See \autoref{fig:eval_config_space} for the prefill-heavy workload's per-batch accuracy.
    }
    \label{fig:per-batch-accuracy-decode}
\end{figure}

\autoref{fig:per-batch-accuracy-prefill} shows the per-batch accuracy on Command-R7B across three attention backends and two GPUs.
Mirroring the Llama-3.1-8B results in \autoref{fig:eval_config_space}, \sys{}Sim predicts per-batch latency within 4.5\% MAPE across all configurations, with accuracy improving at higher token counts as latency stabilizes.
\autoref{fig:per-batch-accuracy-decode} shows the corresponding decode-heavy results on Llama-3.1-8B and Command-R7B; despite higher jitter from smaller token counts, MAPE stays below 3.5\% across all configurations.

\section{Inversion Point Validation}
\label{sec:app-inversion}
\mypara{Experiment setup.}
To validate \sys{}Sim's ability to predict the inversion point across different configurations, we use the same setup as in \autoref{sec:app-per-batch}.

\begin{figure}[H]
  \centering
  \vspace{-10pt}
  \includegraphics[width=0.45\linewidth]{inv-legend.pdf}\\
  \vspace{10pt}
  \begin{subfigure}[t]{0.34\linewidth}
    \centering
    \vspace{-110pt}
    \includegraphics[width=\linewidth]{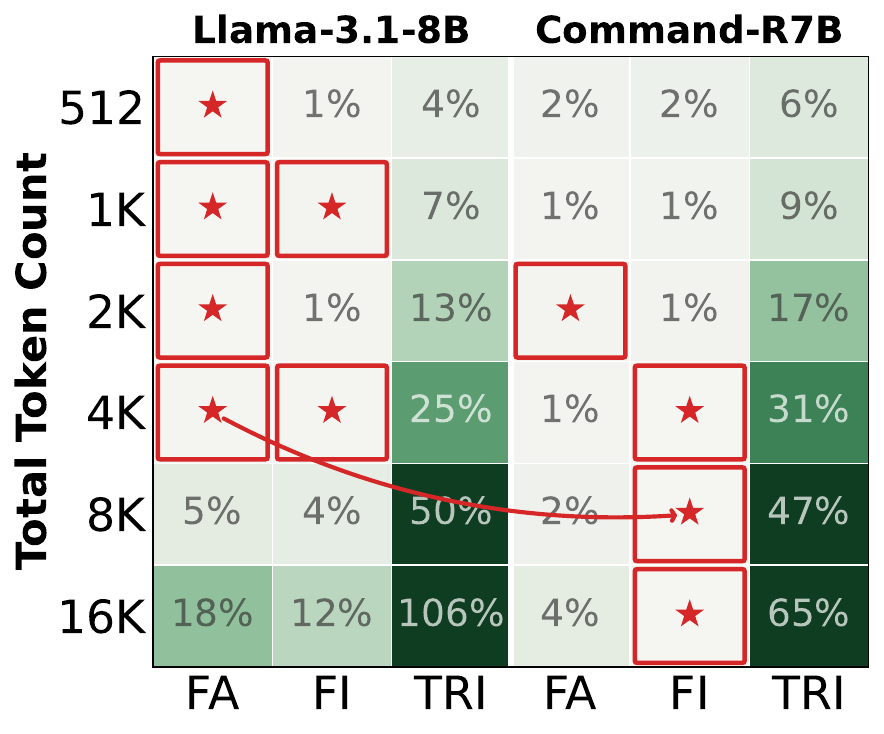}
    \vspace{-13pt}
    \caption{Predicted Latency for Prefill-Heavy Workload}
    \label{fig:inversion_chess_a100:predicted_a100_prefill}
  \end{subfigure}
\hspace{0.08\linewidth}%
  \begin{subfigure}[t]{0.30\linewidth}
    \centering
    \includegraphics[width=\linewidth]{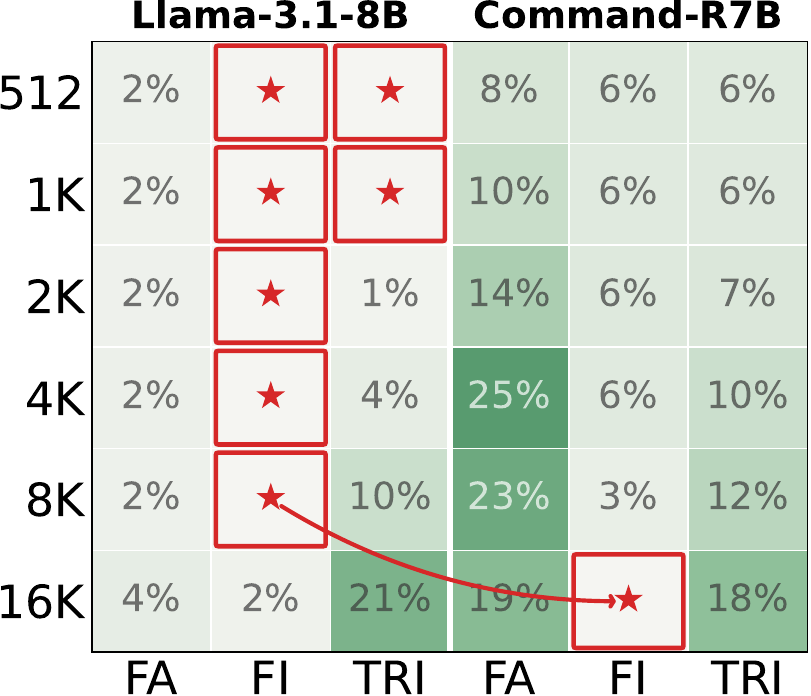}
    \vspace{-13pt}
    \caption{Predicted Latency for Decode-Heavy Workload}
    \label{fig:inversion_chess_a100:predicted_a100_decode}
  \end{subfigure}
  \vspace{-3pt}
  \caption{
    \sys{}Sim correctly predicts the inversion points across different model and attention backend selections on an A100 GPU.
    See \autoref{fig:configuration_space} for the measured inversion points.
    }
    \label{fig:inversion_chess_a100}
\end{figure}

\begin{figure}[H]
  \centering
  \vspace{-10pt}
  \includegraphics[width=0.45\linewidth]{inv-legend.pdf}\\
  \vspace{10pt}
  \begin{subfigure}[t]{0.34\linewidth}
    \centering
    \vspace{-110pt}
    \includegraphics[width=\linewidth]{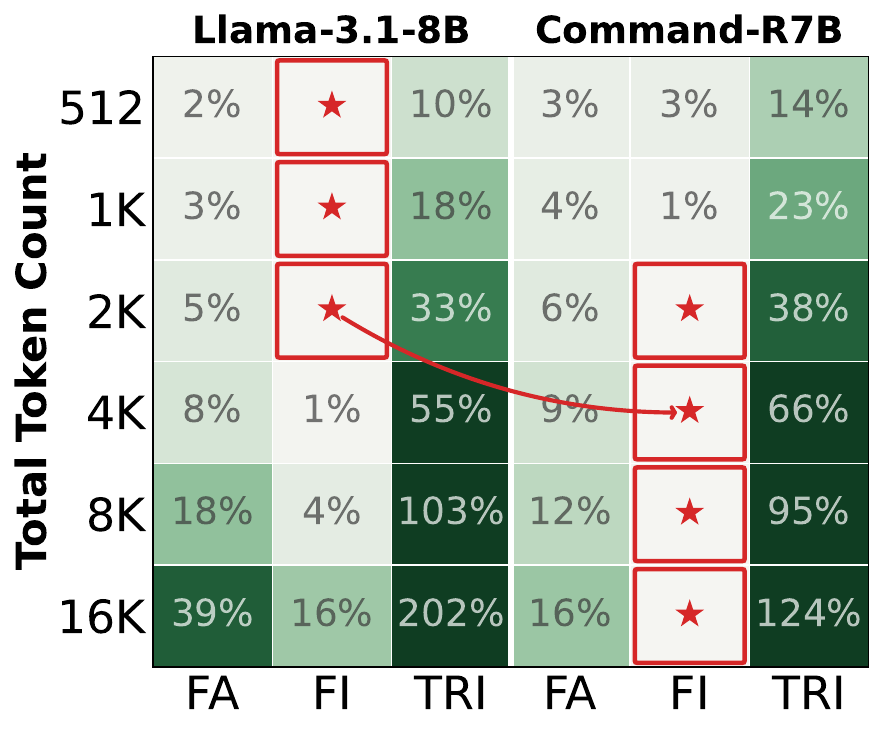}
    \vspace{-13pt}
    \caption{Measured Latency for Prefill-Heavy Workload}
    \label{fig:inversion_chess_h100:measured_h100_prefill}
  \end{subfigure}\hspace{0.08\linewidth}%
  \begin{subfigure}[t]{0.30\linewidth}
    \centering
    \includegraphics[width=\linewidth]{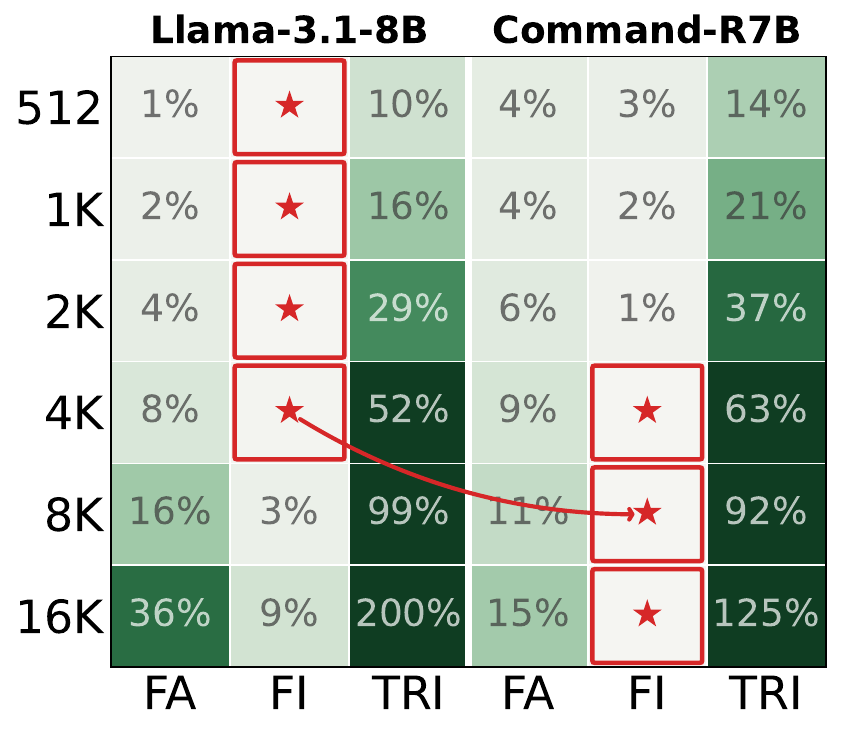}
    \vspace{-13pt}
    \caption{Predicted Latency for Prefill-Heavy Workload}
    \label{fig:inversion_chess_h100:predicted_h100_prefill}
  \end{subfigure}
  \begin{subfigure}[t]{0.34\linewidth}
    \centering
    \includegraphics[width=\linewidth]{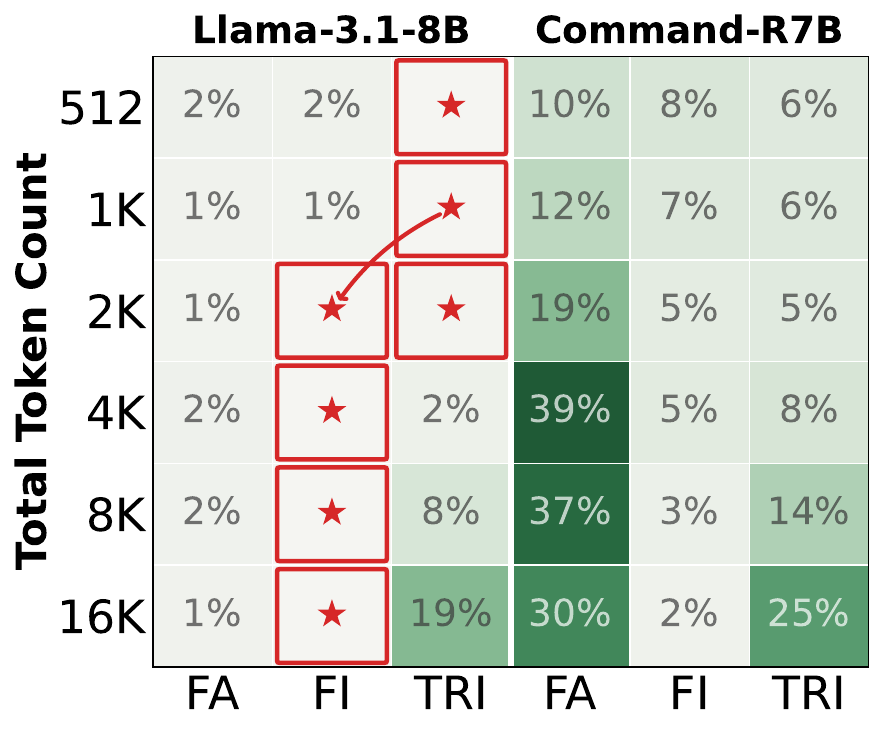}
    \vspace{-13pt}
    \caption{Measured Latency for Decode-Heavy Workload}
    \label{fig:inversion_chess_h100:measured_h100_decode}
  \end{subfigure}
\hspace{0.08\linewidth}%
  \begin{subfigure}[t]{0.32\linewidth}
    \centering
    \includegraphics[width=\linewidth]{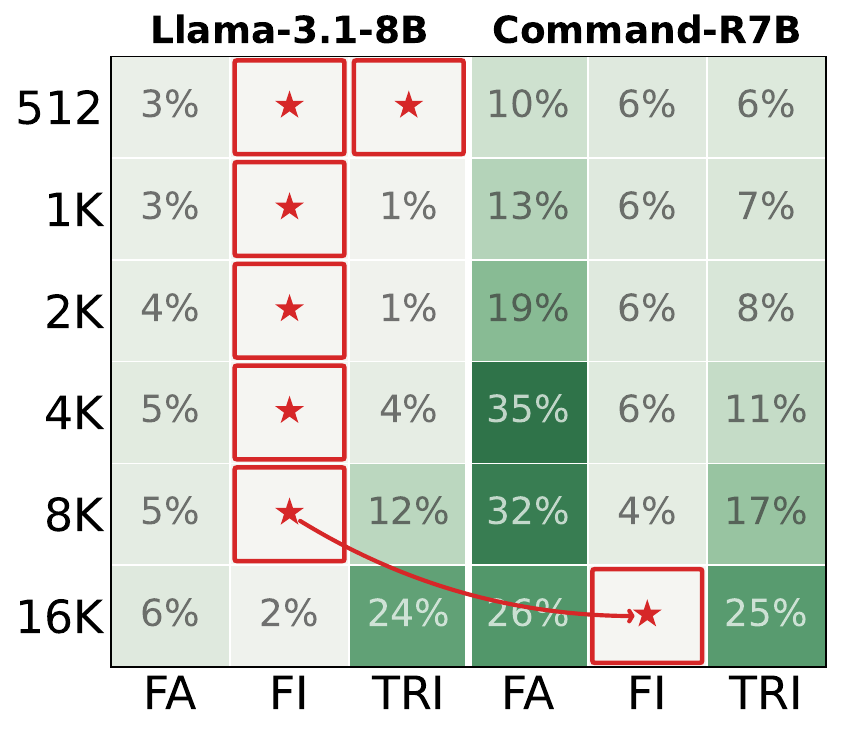}
    \vspace{-13pt}
    \caption{Predicted Latency for Decode-Heavy Workload}
    \label{fig:inversion_chess_h100:predicted_h100_decode}
  \end{subfigure}
  \vspace{-3pt}
  \caption{
    \sys{}Sim correctly predicts the inversion points across different model and attention backend selections on an H100 GPU.
    }
    \label{fig:inversion_chess_h100}
\end{figure}

\sys{}Sim's high per-batch accuracy enables it to correctly predict the inversion points across model and attention backend selections on both A100 (\autoref{fig:inversion_chess_a100}) and H100 (\autoref{fig:inversion_chess_h100}) GPUs.

\section{Simulation Accuracy Under Pipeline, Expert, and Data Parallelism}
\label{sec:app-parallelism}

Every parallelism strategy composes computation and communication latencies, both of which \sys{} already profiles independently; only their composition differs across strategies, which the simulator handles with per-strategy logic.
Supporting a new strategy therefore reduces to simulation engineering rather than profiling changes.

\mypara{Pipeline parallelism (PP).}
PP decomposes into per-shard compute and point-to-point send/receive between $N$ stages: $t_{\text{total}} = t_{\text{compute}} + (N-1)\,t_{\text{p2p}}$.
We extend \sys{}Sim accordingly and evaluate Llama-3.1-8B with PP=2 on two A100 GPUs (\autoref{tab:pp-accuracy}).
TPOT error exceeds TTFT error because decode is more susceptible to timing jitter (\autoref{sec:eval-accuracy}).

\begin{table}[h]
  \centering
  \small
  \setlength{\tabcolsep}{5pt}
  \begin{tabular}{@{}rrrrrrr@{}}
    \toprule
    & \multicolumn{3}{c}{\textbf{TTFT}} & \multicolumn{3}{c}{\textbf{TPOT}} \\
    \cmidrule(lr){2-4} \cmidrule(lr){5-7}
    \textbf{Context} & \textbf{Pred. (ms)} & \textbf{Meas. (ms)} & \textbf{Err. (\%)} & \textbf{Pred. (ms)} & \textbf{Meas. (ms)} & \textbf{Err. (\%)} \\
    \midrule
    2048  & 145.76  & 140.72  & 3.7 & 11.71 & 12.14 & 3.5 \\
    4096  & 299.85  & 295.10  & 1.6 & 11.81 & 12.42 & 4.9 \\
    8192  & 645.86  & 639.41  & 1.0 & 12.19 & 12.98 & 6.1 \\
    16384 & 1475.67 & 1464.12 & 0.8 & 13.00 & 13.94 & 6.7 \\
    \bottomrule
  \end{tabular}
  \vspace{5pt}
  \caption{\sys{}Sim prediction accuracy for Llama-3.1-8B with pipeline parallelism (PP=2) on A100 GPUs.}
  \label{tab:pp-accuracy}
\end{table}

\mypara{Expert parallelism (EP).}
With the per-block (tokens, activated experts) latency tables of \autoref{sec:app-moe}, determining which experts activate, and thus the per-rank load under EP, is the simulator's responsibility.
We extend \sys{}Sim to compose the profiled latencies under a balanced routing algorithm.
For Qwen1.5-MoE-A2.7B at EP=2, TP=2 on NVLink-connected A100 GPUs, the composition matches measured vLLM TTFT within 7.4\% and TPOT within 6.2\% (\autoref{tab:ep-accuracy}).

\begin{table}[h]
  \centering
  \small
  \setlength{\tabcolsep}{5pt}
  \begin{tabular}{@{}rrrrrrr@{}}
    \toprule
    & \multicolumn{3}{c}{\textbf{TTFT}} & \multicolumn{3}{c}{\textbf{TPOT}} \\
    \cmidrule(lr){2-4} \cmidrule(lr){5-7}
    \textbf{Context} & \textbf{Pred. (ms)} & \textbf{Meas. (ms)} & \textbf{Err. (\%)} & \textbf{Pred. (ms)} & \textbf{Meas. (ms)} & \textbf{Err. (\%)} \\
    \midrule
    2048 & 44.8  & 48.3  & 7.4 & 5.76 & 5.42 & 6.2 \\
    4096 & 77.0  & 82.0  & 6.2 & 6.96 & 6.69 & 4.1 \\
    8000 & 146.8 & 157.9 & 7.0 & 9.18 & 8.93 & 2.8 \\
    \bottomrule
  \end{tabular}
  \vspace{5pt}
  \caption{\sys{}Sim prediction accuracy for Qwen1.5-MoE-A2.7B with expert parallelism (EP=2, TP=2) on A100 GPUs.}
  \label{tab:ep-accuracy}
\end{table}

\vspace{-10pt}
\mypara{Data parallelism (DP).}
Model replicas run on different GPUs, each serving separate requests.
Since each replica behaves identically to the single-GPU (or TP shard) case, the per-configuration accuracy results in \autoref{sec:eval-accuracy} already validate per-replica latency; simulating DP requires only spawning $N$ independent simulation instances, with no changes to the profiler or simulator logic.

\mypara{Context parallelism.}
Sequence parallelism is not a standard inference axis; instead, vLLM exposes Decode Context Parallelism (DCP) and an experimental prefill counterpart, which shard the KV cache and attention along the context dimension and reconcile outputs with an all-gather and a reduce-scatter.
\sys{} extends to these in principle by profiling attention at its context-sharded shape (as for TP) and the two collectives per topology (as for TP, PP, and EP); we leave an end-to-end evaluation to future work.

\section{Profiling Overhead: Setup and SGLang Results}
\label{sec:app-profiling-overhead}

\begin{figure}[H]
  \centering
  \begin{minipage}[c]{0.42\linewidth}
    \centering
    \includegraphics[width=\linewidth]{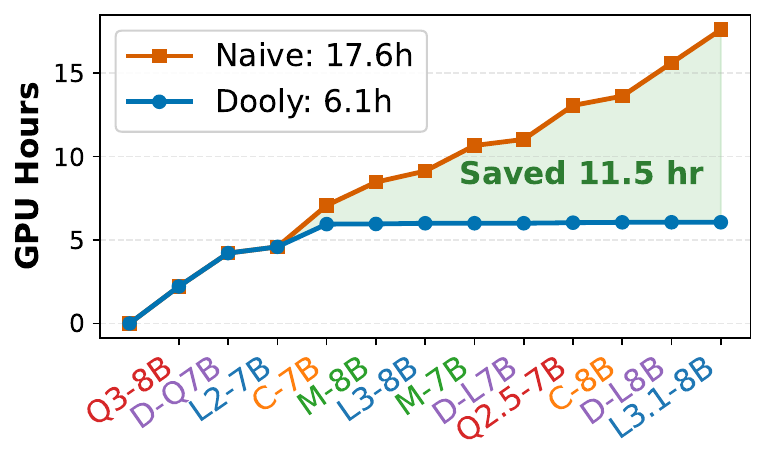}
    \vspace{-20pt}
    \captionof{figure}{
      SGLang counterpart to \autoref{fig:overlap}: deduplication reduces profiling cost by 65.6\% by sharing latency measurements across configurations.
      }
      \label{fig:sglang-overlap}
    \end{minipage}
    \hspace{0.03\linewidth}%
    \begin{minipage}[c]{0.54\linewidth}
      \centering
      \vspace{15pt}
      \vspace{-10pt}
      \footnotesize
      \setlength{\tabcolsep}{1pt}
      \renewcommand{\arraystretch}{0.9}
      \begin{tabular}{@{}llrrrrr@{}}
        \toprule
        \textbf{Group} & \textbf{Variant} & \textbf{N} & \textbf{R} & \textbf{Profile (h)} & \textbf{Saved (h)} & \textbf{Red. (\%)} \\
        \midrule
        \multirow{6}{*}{\textbf{Attention}} & \textit{(aggregate)} & 37 & 26 & 5.28 & 10.35 & 66.2 \\
        & GQA 32/8/128   & 23 & 20 & 2.01 & 8.25 & 80.4 \\
        & GQA 28/4/128   & 6 & 3 & 1.76 & 1.76 & 50.0 \\
        & MHA 32/32/128 & 6 & 3 & 0.34 & 0.34 & 50.0 \\
        & SWA window=4K  & 2 & 0 & 1.17 & 0.00 & 0.0 \\
        \midrule
        \multicolumn{2}{@{}l}{Linear (aten::linear)} & 140 & 121 & 0.59 & 1.01 & 63.0 \\
        \multicolumn{2}{@{}l}{Other operators} & 213 & 175 & 0.19 & 0.18 & 48.4 \\
        \midrule
        \multicolumn{2}{@{}l}{\textbf{Total}} & \textbf{390} & \textbf{322} & \textbf{6.06} & \textbf{11.54} & \textbf{65.6} \\
        \bottomrule
      \end{tabular}
      \captionsetup{type=table}
      \captionof{table}{
        Per-operation profiling savings for SGLang.
        \textbf{N}: number of occurrences; \textbf{R}: number of reuses; \textbf{Profile}: total profiling latency; \textbf{Saved}: total hours saved; \textbf{Red.}: \% reduction.
        }
        \label{tab:sglang-profiling-savings}
        \vspace{5pt}
      \end{minipage}
    \end{figure}

\mypara{Experiment setup.}
To verify the profiling overhead reduction from deduplication, we profile our 12-model corpus across three attention backends and measure the total profiling time with and without deduplication.
For~\autoref{fig:overlap}, we use a 32K prefill chunk size (commonly used in production) and a maximum batch size of 256 (vLLM's default)~\cite{max-num-batched-tokens-rocm, max-num-batched-tokens-vllm, max-num-batched-tokens-qwen, vllm}.
For the SGLang~\cite{sglang} experiment in~\autoref{fig:sglang-overlap}, we keep the same prefill chunk size but use a smaller maximum batch size of 64.

The labels used in \autoref{fig:overlap} and \autoref{fig:sglang-overlap} are defined as follows:
\begin{packeditemize}
    \item \textbf{L3-8B}: \texttt{meta-llama/Meta-Llama-3-8B}
    \item \textbf{L3.1-8B}: \texttt{meta-llama/Llama-3.1-8B}
    \item \textbf{L2-7B}: \texttt{meta-llama/Llama-2-7b-hf}
    \item \textbf{Q2.5-7B}: \texttt{Qwen/Qwen2.5-7B}
    \item \textbf{Q3-8B}: \texttt{Qwen/Qwen3-8B}
    \item \textbf{C-8B}: \texttt{CohereLabs/aya-expanse-8b}
    \item \textbf{C-7B}: \texttt{CohereLabs/c4ai-command-r7b-12-2024}
    \item \textbf{M-7B}: \texttt{mistralai/Mistral-7B-v0.3}
    \item \textbf{M-8B}: \texttt{mistralai/Ministral-8B-Instruct-2410}
    \item \textbf{D-L7B}: \texttt{deepseek-ai/deepseek-llm-7b-base}
    \item \textbf{D-L8B}: \texttt{deepseek-ai/DeepSeek-R1-Distill-Llama-8B}
    \item \textbf{D-Q7B}: \texttt{deepseek-ai/DeepSeek-R1-Distill-Qwen-7B}
\end{packeditemize}

\mypara{Deduplication savings for SGLang.}
SGLang shows a 65.6\% reduction in profiling time through deduplication, with Attention again dominating the total cost (\autoref{tab:sglang-profiling-savings}), consistent with the vLLM results in \autoref{tab:profiling-savings}.
Unlike vLLM, SGLang does not distinguish the interleaved attention kernels, so the Ministral model's profiling cost cannot be amortized.

\vspace{-10pt}
    \vspace{-20pt}
\section{Profiling Savings for Larger and MoE Models}
\label{sec:app-profiling-overhead-extended}

\subsection{Larger Models (13B--72B)}
To test whether deduplication savings hold beyond the 7B--8B corpus, we extend the analysis of \autoref{tab:profiling-savings} to nine models from 13B to 72B parameters, with no changes to \sys{}.
As shown in \autoref{tab:large-model-savings}, deduplication saves 8.21 GPU-hours, a 62.4\% reduction.
Attention again dominates, accounting for 91\% of the savings: four models across two Qwen generations and a DeepSeek release (14B--32B) collapse to one attention signature, and the three largest models (TP=4) collapse to another.
This works because \sys{} profiles tensor-parallel models at per-rank geometry by emulating a single shard (\autoref{sec:tainted_runner}).
At TP=4, the 70B--72B models' 64/8/128 attention becomes 16/2/128 per rank, so compute operations deduplicate against any model sharing that per-rank geometry.
Communication introduced by tensor parallelism is profiled separately and reused across all configurations sharing the same topology and TP degree (\autoref{sec:app-database}).

\begin{table}[h]
  \centering
  \footnotesize
  \setlength{\tabcolsep}{3pt}
  \begin{tabular}{@{}llrrrrr@{}}
    \toprule
    \textbf{Group} & \textbf{Variant} & \textbf{N} & \textbf{R} & \textbf{Profile (h)} & \textbf{Saved (h)} & \textbf{Red. (\%)} \\
    \midrule
    \multirow{5}{*}{\textbf{Attention}} & \textit{(aggregate)} & 30 & 15 & 3.82 & 7.47 & 66.2 \\
    & DeepSeek-Qwen-32B, Qwen2.5-14B/32B, Qwen3-14B & 12 & 9 & 0.99 & 5.54 & 84.8 \\
    & Llama-2-70B, Llama-3.3-70B, Qwen2.5-72B & 9 & 6 & 1.61 & 1.93 & 54.5 \\
    & Llama-2-13B & 3 & 0 & 0.26 & 0.00 & 0.0 \\
    & Gemma-2-27B & 6 & 0 & 0.96 & 0.00 & 0.0 \\
    \midrule
    \multicolumn{2}{@{}l}{Linear (aten::linear)} & 135 & 107 & 0.98 & 0.63 & 39.1 \\
    \multicolumn{2}{@{}l}{Other operators} & 252 & 207 & 0.14 & 0.11 & 43.7 \\
    \midrule
    \multicolumn{2}{@{}l}{\textbf{Total}} & \textbf{417} & \textbf{329} & \textbf{4.94} & \textbf{8.21} & \textbf{62.4} \\
    \bottomrule
  \end{tabular}
  \vspace{5pt}
  \caption{
    Per-operation profiling savings for nine models from 13B to 72B parameters.
    Columns follow \autoref{tab:profiling-savings}.
  }
  \label{tab:large-model-savings}
\end{table}

\subsection{Mixture-of-Experts Models}
\label{sec:app-moe}
\sys{} supports MoE layers by adding the number of activated experts as a sweep dimension.
A hook on vLLM's \texttt{FusedMoE} module, the common entry point for every MoE architecture in vLLM, forces exactly the target number of distinct experts to fire, producing a (tokens, activated experts) latency table per MoE block that is independent of the model's learned routing.

\mypara{Routing independence.}
Routing affects MoE latency only through the token count and the number of activated experts: since experts share identical shapes, latency is independent of which experts fire or of token values.
Gate-router latency varies by at most 1.4\% across random inputs at a fixed shape.
Consequently, \texttt{FusedMoE} signatures generalize across architectures: \autoref{tab:moe-fusedmoe} shows that matching signatures yield nearly identical latency for DeepSeek-V2-Lite (DVL) and DeepSeek-MoE-16B (DM), despite different model weights.

\begin{table}[h]
  \centering
  \small
  \setlength{\tabcolsep}{6pt}
  \begin{tabular}{@{}rrrrr@{}}
    \toprule
    \textbf{\# Tokens} & \textbf{\# Activated Experts} & \textbf{DVL (ms)} & \textbf{DM (ms)} & \textbf{Diff. (\%)} \\
    \midrule
    512  & 16 & 0.57 & 0.57 & 0.5 \\
    512  & 64 & 0.95 & 0.94 & 0.4 \\
    1024 & 16 & 1.01 & 1.00 & 0.2 \\
    2048 & 64 & 2.38 & 2.39 & 0.2 \\
    \bottomrule
  \end{tabular}
  \vspace{5pt}
  \caption{Profiled \texttt{FusedMoE} latency for DeepSeek-V2-Lite (DVL) and DeepSeek-MoE-16B (DM).}
  \label{tab:moe-fusedmoe}
\end{table}

\mypara{Deduplication savings.}
\autoref{tab:moe-savings} extends the deduplication analysis of \autoref{tab:profiling-savings} to a corpus of dense and MoE models.
\sys{} saves 61.4\% of GPU-hours overall and 72.8\% for MoE operations, with no model-specific implementation.

\begin{table}[h]
  \centering
  \footnotesize
  \setlength{\tabcolsep}{3pt}
  \begin{tabular}{@{}llrrrrr@{}}
    \toprule
    \textbf{Group} & \textbf{Variant} & \textbf{N} & \textbf{R} & \textbf{Profile (h)} & \textbf{Saved (h)} & \textbf{Red. (\%)} \\
    \midrule
    \multirow{4}{*}{\textbf{Attention}} & \textit{(aggregate)} & 16 & 2.3 & 1.59 & 2.47 & 60.8 \\
    & Llama-3.1-8B, Mistral-7B, Mixtral-8$\times$7B & 9 & 3.0 & 0.93 & 1.99 & 68.2 \\
    & Qwen1.5-MoE, DM & 6 & 2.0 & 0.27 & 0.48 & 63.5 \\
    & DVL (MLA) & 1 & 1.0 & 0.39 & 0.00 & 0.0 \\
    \midrule
    \multirow{4}{*}{\textbf{FusedMoE}} & \textit{(aggregate)} & 10 & 3.3 & 0.17 & 0.47 & 72.8 \\
    & Mixtral & 3 & 3.0 & 0.14 & 0.28 & 66.7 \\
    & Qwen & 3 & 3.0 & 0.02 & 0.11 & 85.7 \\
    & DVL, DM & 4 & 4.0 & 0.02 & 0.08 & 81.6 \\
    \midrule
    \multicolumn{2}{@{}l}{Linear (aten::linear)} & 72 & 4.5 & 0.10 & 0.09 & 47.3 \\
    \multicolumn{2}{@{}l}{Other operators} & 138 & 4.5 & 0.05 & 0.01 & 21.0 \\
    \midrule
    \multicolumn{2}{@{}l}{\textbf{Total}} & \textbf{236} & \textbf{4.1} & \textbf{1.91} & \textbf{3.03} & \textbf{61.4} \\
    \bottomrule
  \end{tabular}
  \vspace{5pt}
  \caption{
    Per-operation profiling savings across dense and MoE models.
    DVL: DeepSeek-V2-Lite; DM: DeepSeek-MoE-16B.
    Columns follow \autoref{tab:profiling-savings}.
  }
  \label{tab:moe-savings}
\end{table}

\section{Accuracy Cost of Architecture-Level Deduplication}
\label{sec:app-baseline-accuracy}

\begin{table}[h]
  \centering
  \small
  \setlength{\tabcolsep}{4pt}
  \begin{tabular}{@{}lrrrrr@{}}
    \toprule
    \textbf{Context} & \textbf{Measured (ms)} & \textbf{\sys{} Pred. (ms)} & \textbf{HMD Pred. (ms)} & \textbf{\sys{} Err. (\%)} & \textbf{HMD Err. (\%)} \\
    \midrule
    2K  & 153.24  & 159.5  & 159.4  & 4.1 & 4.0 \\
    4K  & 308.36  & 323.0  & 323.2  & 4.8 & 4.8 \\
    8K  & 628.62  & 666.1  & 688.8  & 6.0 & 9.6 \\
    16K & 1310.73 & 1388.0 & 1556.8 & 5.9 & 18.8 \\
    32K & 2788.10 & 2921.9 & 3827.5 & 4.8 & 37.3 \\
    \bottomrule
  \end{tabular}
  \vspace{5pt}
  \caption{TTFT prediction for Command-R7B under \sys{} and HF-Model-Diff (HMD) profiles.}
  \label{tab:ref-hmd-accuracy}
\end{table}

HF-Model-Diff (HMD) deduplicates profiles using only the model architecture parameters in the HuggingFace configuration file.
Command-R7B and Aya-Expanse-8B agree on all of these fields, so HMD reuses Aya-Expanse-8B's profile for Command-R7B.
However, Command-R7B interleaves sliding-window attention, whereas Aya-Expanse-8B uses full attention.
The two attention types call different kernels, as determined by the module-instance state.
This state, however, is invisible from the configuration parameters alone, so Command-R7B's sliding-window attention is never profiled under HMD.
\autoref{tab:ref-hmd-accuracy} shows that HMD's TTFT error grows with context length, from 4.0\% at 2K to 37.3\% at 32K, while \sys{}'s error stays between 4.1\% and 6.0\%.

\section{Ablation: Module-Instance State in Signatures}
\label{sec:app-hashing-ablation}

A naive alternative to \sys{}'s signatures hashes operations by module name, treating every instance of a class as identical.
This breaks for models with heterogeneous layers.
Command-R7B interleaves sliding-window attention (24 layers) and full attention (8 layers), yet all 32 layers are instances of the same attention class with identical tensor shapes and configuration dimensions, so a name-based key merges them.
Adding the dispatched kernel symbol to the key does not help: on the FlashAttention and Triton backends, both attention types dispatch the same kernel, differing only in a runtime argument (the window size).
\sys{} resolves this by folding the module instance's runtime state (here, the sliding-window setting) into the signature.

Name-based hashing introduces no error for Llama-3.1-8B, whose 32 attention layers are homogeneous, but overestimates Command-R7B's attention latency by up to 3.66$\times$ at its 132K-token maximum context: the 24 sliding-window layers are never profiled independently and each inherits the higher latency of the full-attention layers it is merged with.

\section{Profiling Granularity: Accuracy vs.\ Deduplication}
\label{sec:app-granularity}

\mypara{Measurement procedure.}
Given a callable (an operation or a module), \sys{} profiles it with CUDA Graphs: after GPU warm-up, a graph captures 10 runs of the callable, repeated for 20 captures.
Each capture is timed with CUDA events and averaged over its 10 runs, and \sys{} reports the median of the 20 averages.
Because CUDA Graphs execute kernels back to back regardless of Python-level call structure and CUDA events time only kernel execution, module- and operation-level callables that invoke the same kernels yield the same latency.

\begin{table}[h]
  \centering
  \small
  \setlength{\tabcolsep}{6pt}
  \begin{tabular}{@{}rrrrrr@{}}
    \toprule
    \textbf{Tokens} & \textbf{Module ($\mu$s)} & \textbf{Op ($\mu$s)} & \textbf{Abs. gap ($\mu$s)} & \textbf{Rel. gap (\%)} & \textbf{CoV (\%)} \\
    \midrule
    512   & 347.49  & 348.44  & $-$0.96 & $-$0.27 & 0.90 \\
    2048  & 1118.21 & 1120.69 & $-$2.48 & $-$0.22 & 0.49 \\
    8192  & 4309.24 & 4306.06 & 3.18    & 0.07    & 0.27 \\
    16384 & 8519.24 & 8528.62 & $-$9.37 & $-$0.11 & 0.12 \\
    \bottomrule
  \end{tabular}
  \vspace{5pt}
  \caption{\texttt{RowParallelLinear} (module) vs.\ \texttt{aten::linear} (operation) latency (in: 4096, out: 14336).}
  \label{tab:granularity-linear}
\end{table}

\mypara{Dispatcher-level operation vs.\ parent module.}
We compare \texttt{aten::linear}, captured via the PyTorch dispatcher, against its parent module \texttt{RowParallelLinear} using Llama-3.1-8B's dimensions ($[X, 4096] \times [4096, 14336]$, $X$ swept from 512 to 16384 tokens).
Both paths invoke the same cuBLAS kernel; \autoref{tab:granularity-linear} shows relative differences under 1\% with comparable jitter.

\begin{table}[h]
  \centering
  \small
  \setlength{\tabcolsep}{5pt}
  \begin{tabular}{@{}lrrrrrr@{}}
    \toprule
    \textbf{Implementation} & \textbf{Tokens} & \textbf{Module ($\mu$s)} & \textbf{Op ($\mu$s)} & \textbf{Abs. gap ($\mu$s)} & \textbf{Rel. gap (\%)} & \textbf{CoV (\%)} \\
    \midrule
    \multirow{4}{*}{Gemma RMSNorm}
      & 512   & 5.38   & 5.38   & 0.0007  & 0.013   & 0.42 \\
      & 2048  & 21.79  & 21.80  & $-$0.01 & $-$0.027 & 0.33 \\
      & 8192  & 83.59  & 83.62  & $-$0.03 & $-$0.033 & 0.13 \\
      & 16384 & 161.64 & 161.66 & $-$0.03 & $-$0.017 & 0.07 \\
    \midrule
    \multirow{4}{*}{Cohere LayerNorm}
      & 512   & 6.36   & 6.35   & 0.003   & 0.06    & 0.15 \\
      & 2048  & 22.56  & 22.55  & 0.013   & 0.06    & 0.29 \\
      & 8192  & 85.05  & 85.03  & 0.0157  & 0.02    & 0.14 \\
      & 16384 & 165.76 & 165.81 & $-$0.05 & $-$0.03 & 0.27 \\
    \bottomrule
  \end{tabular}
  \vspace{5pt}
  \caption{Triton-fused custom modules vs.\ their \texttt{forward\_impl} operations.}
  \label{tab:granularity-triton}
\end{table}

\vspace{-10pt}
\mypara{Triton-fused custom kernels.}
We repeat the comparison for \texttt{torch.compile}-fused Triton normalization kernels that bypass the dispatcher entirely (Gemma RMSNorm and Cohere LayerNorm), comparing each module against its \texttt{forward\_impl} operation.
\autoref{tab:granularity-triton} shows relative gaps under 0.1\%.

\mypara{Cost of coarser granularity.}
The cost of falling back to module-level profiling is a lost deduplication opportunity, not a loss of accuracy.
For instance, Mamba-Codestral-7B-v0.1's \texttt{MambaMixer2} module calls two linear operations, an RMS normalization, and the \texttt{mamba\_mixer2} operation.
Because \texttt{mamba\_mixer2} is context-dependent, the Operation Set Finder profiles the enclosing \texttt{MambaMixer2} module and absorbs its sibling operations (\autoref{sec:testing_runnability}); the linear operations, which could otherwise be shared across configurations, are no longer deduplicated.
More generally, a standalone operation is absorbed only when a context-dependent operation forces a fallback to an enclosing module that also contains it.
This does not occur for attention or MoE, whose context-dependent operations resolve at a tight submodule separate from the standalone operations.



\makeatletter
\@ifpackagewith{neurips_2026}{preprint}{}{
  \clearpage
  \section*{NeurIPS Paper Checklist}

\begin{enumerate}

\item {\bf Claims}
    \item[] Question: Do the main claims made in the abstract and introduction accurately reflect the paper's contributions and scope?
    \item[] Answer: \answerYes{} 
    \item[] Justification: We explain the problems mentioned in the abstract and introduction in \autoref{sec:background} and explain our solutions in~\autoref{sec:overview}. 
    Further detail is provided in~\autoref{sec:tainted_runner},\autoref{sec:op_set_finder},\autoref{sec:duplication_aware_profiler}.

\item {\bf Limitations}
    \item[] Question: Does the paper discuss the limitations of the work performed by the authors?
    \item[] Answer: \answerYes{} 
    \item[] Justification: We discuss the limitations of our work in \autoref{sec:limitations}.

\item {\bf Theory assumptions and proofs}
    \item[] Question: For each theoretical result, does the paper provide the full set of assumptions and a complete (and correct) proof?
    \item[] Answer: \answerNA{} 
    \item[] Justification: \answerNA{}

    \item {\bf Experimental result reproducibility}
    \item[] Question: Does the paper fully disclose all the information needed to reproduce the main experimental results of the paper to the extent that it affects the main claims and/or conclusions of the paper (regardless of whether the code and data are provided or not)?
    \item[] Answer: \answerYes{} 
    \item[] Justification: 
        We provide detailed information about the experiments including the models used, the public dataset used, and the configurations used for the experiments.
        We also provide the RPS value used during the streaming experiments.
        The code will be publicly released upon publication.

\item {\bf Open access to data and code}
    \item[] Question: Does the paper provide open access to the data and code, with sufficient instructions to faithfully reproduce the main experimental results, as described in supplemental material?
    \item[] Answer: \answerYes{} 
    \item[] Justification: 
        We used open datasets for experiments.
        We will release the source code used for our experiments upon publication.

\item {\bf Experimental setting/details}
    \item[] Question: Does the paper specify all the training and test details (e.g., data splits, hyperparameters, how they were chosen, type of optimizer) necessary to understand the results?
    \item[] Answer: \answerYes{} 
    \item[] Justification: 
        We provide details about the experiment set-up in \autoref{sec:app-sim-accuracy} and \autoref{sec:app-per-batch}.

\item {\bf Experiment statistical significance}
    \item[] Question: Does the paper report error bars suitably and correctly defined or other appropriate information about the statistical significance of the experiments?
    \item[] Answer: \answerYes{} 
    \item[] Justification:
        We report the MAE across all percentiles of the CDF in the streaming experiments.
        We also report the error bars in the per-batch experiments for all figures.
        Note that the error bars are not visible in some configurations due to the small variance of latency across runs that have been measured after multiple warm up runs. 

\item {\bf Experiments compute resources}
    \item[] Question: For each experiment, does the paper provide sufficient information on the computer resources (type of compute workers, memory, time of execution) needed to reproduce the experiments?
    \item[] Answer: \answerYes{} 
    \item[] Justification: 
        We report the specific GPU used for all experiments. 
    
\item {\bf Code of ethics}
    \item[] Question: Does the research conducted in the paper conform, in every respect, with the NeurIPS Code of Ethics \url{https://neurips.cc/public/EthicsGuidelines}?
    \item[] Answer: \answerYes{} 
    \item[] Justification: 
        The application of this work to LLM inference simulations does not have negative societal impacts or ethical concerns. 
        Our method does not have negative societal impacts, as we use publicly released data and model checkpoints. 

\item {\bf Broader impacts}
    \item[] Question: Does the paper discuss both potential positive societal impacts and negative societal impacts of the work performed?
    \item[] Answer: \answerNA{} 
    \item[] Justification:
        \sys{} is a profiling tool for LLM inference systems with no direct path to negative societal applications; we note this in \autoref{sec:concl}.
    
\item {\bf Safeguards}
    \item[] Question: Does the paper describe safeguards that have been put in place for responsible release of data or models that have a high risk for misuse (e.g., pre-trained language models, image generators, or scraped datasets)?
    \item[] Answer: \answerNA{} 
    \item[] Justification: \answerNA{}

\item {\bf Licenses for existing assets}
    \item[] Question: Are the creators or original owners of assets (e.g., code, data, models), used in the paper, properly credited and are the license and terms of use explicitly mentioned and properly respected?
    \item[] Answer: \answerYes{} 
    \item[] Justification: We cite the original paper\slash website\slash license used in our experiments.

\item {\bf New assets}
    \item[] Question: Are new assets introduced in the paper well documented and is the documentation provided alongside the assets?
    \item[] Answer: \answerNA{} 
    \item[] Justification: This paper does not release new assets.

\item {\bf Crowdsourcing and research with human subjects}
    \item[] Question: For crowdsourcing experiments and research with human subjects, does the paper include the full text of instructions given to participants and screenshots, if applicable, as well as details about compensation (if any)? 
    \item[] Answer: \answerNA{} 
    \item[] Justification: \answerNA{}

\item {\bf Institutional review board (IRB) approvals or equivalent for research with human subjects}
    \item[] Question: Does the paper describe potential risks incurred by study participants, whether such risks were disclosed to the subjects, and whether Institutional Review Board (IRB) approvals (or an equivalent approval/review based on the requirements of your country or institution) were obtained?
    \item[] Answer: \answerNA{} 
    \item[] Justification: \answerNA{}

\item {\bf Declaration of LLM usage}
    \item[] Question: Does the paper describe the usage of LLMs if it is an important, original, or non-standard component of the core methods in this research? Note that if the LLM is used only for writing, editing, or formatting purposes and does \emph{not} impact the core methodology, scientific rigor, or originality of the research, declaration is not required.
    \item[] Answer: \answerNA{} 
    \item[] Justification: \answerNA{}

\end{enumerate}
}
\end{document}